\documentclass[aps,pra,reprint,showpacs,notitlepage,superscriptaddress,onecolumn]{revtex4-2}
\usepackage{amssymb}
\usepackage{mathrsfs}
\usepackage{mathtools}
\usepackage{amsfonts}
\usepackage{graphicx}
\usepackage{amsmath}
\usepackage{dcolumn}
\usepackage{comment}
\usepackage{xcolor}
\usepackage{subfigure}
\usepackage{booktabs}
\usepackage{epsfig}
\usepackage{soul}
\usepackage[normalem]{ulem}
\usepackage[colorlinks,linkcolor=blue,citecolor=blue,hyperindex,bookmarks=false,pdfstartview=FitH,urlcolor=blue]{hyperref}
\usepackage{orcidlink}

\providecommand{\U}[1]{\protect\rule{.1in}{.1in}}
\definecolor{Blue}{rgb}{0.00,0.00,0.80}
\definecolor{Red}{rgb}{0.80,0.00,0.00}

\newcommand{\ignore}[1]{}
\newcommand{\figpanel}[2]{\hyperref[#1]{\ref*{#1}(#2)}}

\begin{document}

\title{Nonlinear quantum optomechanics in a Fano-mirror microcavity system}

\author{Lei Du \orcidlink{0000-0003-0641-440X}}
\email{lei.du@chalmers.se}
\affiliation{Department of Microtechnology and Nanoscience (MC2), Chalmers University of Technology, 412 96 Gothenburg, Sweden}
\author{Juliette Monsel \orcidlink{0000-0002-4965-6794}}
\affiliation{Department of Microtechnology and Nanoscience (MC2), Chalmers University of Technology, 412 96 Gothenburg, Sweden}
\author{Witlef Wieczorek \orcidlink{0000-0003-1847-053X}}
\affiliation{Department of Microtechnology and Nanoscience (MC2), Chalmers University of Technology, 412 96 Gothenburg, Sweden}
\author{Janine Splettstoesser \orcidlink{0000-0003-1078-9490}}
\affiliation{Department of Microtechnology and Nanoscience (MC2), Chalmers University of Technology, 412 96 Gothenburg, Sweden}

\begin{abstract}
We study a Fano-mirror optomechanical system in the quantum nonlinear regime. In this system, two strongly lossy optical modes hybridize through both coherent and dissipative couplings to form an effective optical mode with a drastically reduced linewidth. This linewidth reduction enables the system to access the single-photon strong-coupling and sideband-resolved regimes simultaneously. We formulate the system dynamics using an effective master-equation approach and benchmark it against quantum Langevin and dressed-state master-equation descriptions. With experimentally realistic parameters, we predict clear quantum signatures, including photon blockade and the generation of mechanical cat states. Our work establishes the Fano-mirror architecture as a promising platform for harnessing single-photon optomechanical nonlinearities for quantum state engineering under achievable experimental conditions.
\end{abstract}

\maketitle

\section{Introduction}\label{SecIntro}

Cavity optomechanical systems constitute a powerful platform for precision sensing and quantum control of mechanical motion~\cite{OM2014RMP,OMQT,OConnell2010,OMmeasure1,OMmeasure2,OMmeasure3}. However, since the native radiation–pressure interaction is often weak, most experiments use strong laser drives to boost the effective optomechanical interaction~\cite{Teufel2011Nature,Groblacher2009Nature}. Although this driven regime can significantly enhance the interaction strength, it linearizes the system dynamics \cite{OM2014RMP} and thus obscures the intrinsic optomechanical nonlinearity. While Kerr-type cavity nonlinearities can facilitate certain tasks (e.g., cooling or amplification) in the sideband-unresolved regime~\cite{Clerk2011Kerr,Bothner2022Kerr,Metelmann2023Kerr,Metelmann2025Kerr}, they normally rely on additional nonlinear elements and often involve a tradeoff between the strong driving needed to exploit the Kerr response and the onset of multistability. This has therefore motivated significant interest in accessing the \emph{intrinsic} single-photon nonlinearity in cavity optomechanics, which would enable deterministic generation of nonclassical states~\cite{Bose1997PRA,Marshall2003PRL,RablPB2011,SVCR2,Garziano2015PRA,Hauer2023prl} and open up sensing protocols that are inaccessible in the linear regime~\cite{Pinard1995PRA,Qvarfort2018NC}.  Microcavity implementations can enhance the single-photon optomechanical coupling rate \(g_{0}\) by shrinking the cavity length (and thus the optical mode volume), but they suffer from large optical loss. In state-of-the-art devices, the optical decay rate $\kappa$ usually exceeds $g_{0}$ by two orders of magnitude~\cite{Teufel2011Nature,Chan2011Nature,Leijssen2017NC}. This makes it challenging to simultaneously achieve single-photon strong coupling and sideband resolution, and thereby makes nonlinear optomechanical effects difficult to resolve and exploit.

\begin{figure}
\centering
\includegraphics[width = 0.6\linewidth]{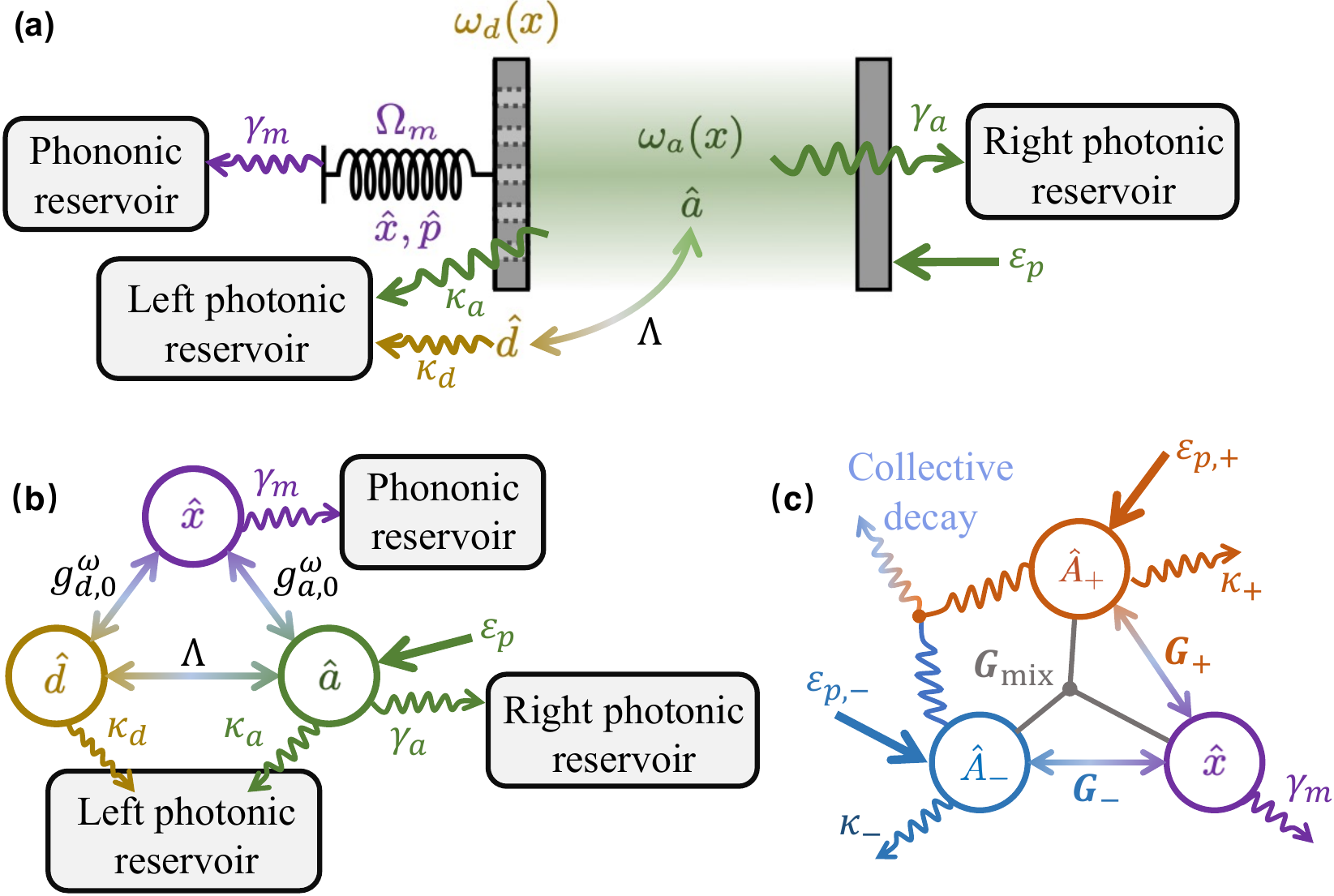}
\caption{(a) Schematic illustration of the Fano-mirror optomechanical system.
(b) Diagram of the original Fano-mirror optomechanical setup, including the interactions among the two optical modes and the mechanical mode, see Sec.~\ref{SecModel}.
(c) Diagram of the system in the optical normal-mode basis, with all effective dissipation and interaction channels indicated, see Sec.~\ref{SecNormal}.}
\label{Model}
\end{figure}

A recently developed  architecture~\cite{FanoM,WWoe2023,Juliette2023pra,MitraOE2024,DL2025Fano} has emerged as a promising candidate to overcome this roadblock. It comprises a microcavity formed by a suspended photonic crystal (PhC) membrane and a highly reflective, frequency-independent mirror, see Fig.~\figpanel{Model}{a}. The PhC membrane supports both a mechanical vibrational mode and a localized optical mode. The standing-wave cavity mode and the localized optical mode hybridize through coherent coupling and, crucially, through a \emph{strong dissipative coupling} mediated by a shared photonic reservoir~\cite{FanoM}. This hybridization yields bright and dark optical normal modes; in the dark mode, the effective optical loss is strongly suppressed while the optomechanical coupling remains appreciable. As a result, an effective sideband-resolved regime, where the optical linewidth becomes smaller than the mechanical frequency, can be reached~\cite{Juliette2023pra,DL2025Fano}. The optical dissipative coupling, typically absent in conventional coupled-mode optomechanical systems~\cite{2cavity1,2cavity2,2cavity3,YCLiu2015pra}, plays a crucial role in significantly reducing the optical loss and thereby enhancing the sideband resolution. 

Previous theoretical studies of such optomechanical systems (hereafter referred to as ``Fano-mirror optomechanical systems'', due to the Fano-type interference between the two optical modes~\cite{FanoM}) have predominantly focused on the \emph{linear} regime~\cite{FanoM, Juliette2023pra, DL2025Fano}, where the single-photon optomechanical coupling strengths are much smaller than the mechanical frequency. However, recent experiments~\cite{WWoe2023} have demonstrated that these couplings can become comparable to, or even exceed, the mechanical frequency. Motivated by these advances, we go beyond the linear approximation and investigate this architecture in the quantum \emph{nonlinear} regime. Thanks to the nontrivial hybridization between the cavity and localized modes in the presence of the optical dissipative coupling, we identify an \emph{effective} regime in which both single-photon strong coupling and sideband resolution are simultaneously guaranteed, even when the linewidths of these bare optical modes exceed the mechanical frequency by more than five orders of magnitude. This regime enables the observation of both optical and mechanical nonclassical effects, including photon blockade in the dark optical normal mode and deterministic generation of mechanical cat states under different driving protocols. These results are shown to be robust against dissipative contributions of the optomechanical interaction as well as the dressed-state corrections to the decoherence channels. Since the linearization approach previously used in theoretical studies for this system~\cite{FanoM, Juliette2023pra, DL2025Fano} ceases to be adequate in the nonlinear regime, we employ an effective master equation in the optical normal-mode basis, which accurately captures the coherent and dissipative hybridization of the optical modes while keeping the optomechanical interaction fully nonlinear. In the relevant parameter regime, we explicitly benchmark this description against the Langevin-equation~\cite{Juliette2023pra,DL2025Fano} and dressed-state master-equation~\cite{HuDSME} approaches.


\section{Methods}

\subsection{Overview of the Fano-mirror optomechanical setup}\label{SecModel}

We first briefly revisit the Fano-mirror optomechanical setup, which has been theoretically explored~\cite{FanoM,Juliette2023pra,DL2025Fano} and experimentally realized~\cite{WWoe2023,MitraOE2024}. As shown in Fig.~\figpanel{Model}{a}, such a microcavity optomechanical system consists of a normal mirror (e.g., a distributed Bragg reflector) and a suspended PhC mirror. The left PhC mirror supports both a mechanical vibrational mode with frequency $\Omega_{m}$ (characterized by an out-of-plane, dimensionless displacement $x$ and momentum $p$) and a localized optical mode $d$ with frequency $\omega_{d}$. Hereafter, we refer to this localized optical mode as the ``Fano mode''. The intracavity field is described by a cavity mode $a$ with resonance frequency $\omega_{a}$. The two optical modes, i.e., the cavity and Fano modes, are strongly coupled via a coherent interaction at rate $\Lambda$, which originates from the spatial overlap of their electric fields. As a result, the bare mechanical and optical Hamiltonians read ($\hbar=1$ in this paper)
\begin{subequations}
\begin{eqnarray}
\hat{H}_{\mathrm{mec}} &=& \Omega_{m}\hat{b}^{\dag}\hat{b}, \label{baremec}\\
\hat{H}_\text{opt} &=& \omega_a \hat{a}^\dagger \hat{a} + \omega_d \hat{d}^\dagger \hat{d} + \Lambda (\hat{a}^\dagger \hat{d} + \hat{d}^\dagger \hat{a}),
\label{bareopt}
\end{eqnarray}
\end{subequations}
where $\hat{b}=(\hat{b}^{\dag})^{\dag}=(\hat{x}+i\hat{p})/\sqrt{2}$ is the annihilation operator of the mechanical mode, and similarly for the optical modes. Note that the two optical modes are inherently coupled to a common reservoir (i.e., the left photonic reservoir in Fig.~\ref{Model}), which, as will be shown later, yields \emph{collective dissipation} terms in the master equation. This collective dissipation enables nontrivial hybridization effects, and constitutes one of the core ingredients of the present work.

Furthermore, both optical modes interact dispersively with the mechanical mode, with the corresponding Hamiltonian given by
\begin{equation}
\hat{H}_{\text{om}}=-\left(g_{a,0}^{\omega}\hat{a}^\dag\hat{a} + g_{d,0}^{\omega}\hat{d}^\dag\hat{d}\right)\left(\hat{b} + \hat{b}^{\dag}\right),
\label{Hom}
\end{equation}
where $g_{a,0}^{\omega}$ and $g_{d,0}^{\omega}$ are the respective single-photon optomechanical coupling strengths. Specifically, the dispersive coupling between the cavity mode and the mechanical motion $x$ arises from radiation pressure, while that between the Fano and mechanical modes can originate from strain-induced deformations of the PhC mirror, which alter its optical properties. The coupling strengths $g_{a,0}^{\omega}$ and $g_{d,0}^{\omega}$  are defined through the linear expansion
\[
\omega_{a(d)}(x)\simeq \omega_{a(d)}(0)+\left.\frac{\partial \omega_{a(d)}}{\partial x}\right|_{x=0}x
\equiv \omega_{a(d)}(0)-\sqrt{2}\,g_{a(d),0}^{\omega}x.
\]
The explicit minus sign in Eq.~(\ref{Hom}) fixes this convention, but does not imply that the two coupling strengths must have the same sign. With this convention, for the Fabry--P\'{e}rot-like cavity mode, a positive mechanical displacement corresponding to cavity elongation gives $g_{a,0}^{\omega}>0$. By contrast, the Fano-mode coupling typically has the opposite sign (i.e., $g_{d,0}^{\omega}<0$) in experiments~\cite{WWoe2023}. 

Taken together, the total Hamiltonian of the Fano-mirror optomechanical setup, in the absence of external drivings, is given by $\hat{H}_{\mathrm{tot}}=\hat{H}_{\mathrm{opt}}+\hat{H}_{\mathrm{mec}}+\hat{H}_{\mathrm{om}}$. In Fig.~\figpanel{Model}{b}, we provide an intuitive diagram of this three-mode setup, which incorporates the associated interaction and dissipation channels. Specifically, the cavity mode decays through the left PhC and right normal mirrors at rates $\kappa_{a}$ and $\gamma_{a}$, respectively, while the Fano mode decays into the left photonic reservoir at rate $\kappa_{d}$. We also assume a damping rate $\gamma_{m}$ for the mechanical mode. 

In a microcavity geometry, the mechanical displacement $x$ can also considerably modulate the decay rates of both the cavity and Fano modes, leading to \emph{dissipative optomechanical interactions}~\cite{ClerkDissipativeOMcoupling2009,WWoe2023,Juliette2023pra}. To better elucidate the underlying physics, we initially neglect these dissipative contributions. 
Such contributions can be incorporated into the master equation by introducing \emph{position-dependent decay rates} for the optical modes. In Sec.~\ref{SecDissipative}, we will show that, within the parameter regimes considered in this paper, the dissipative optomechanical couplings have a negligible effect on the system dynamics.

\subsection{Optical normal modes}\label{SecNormal}

In this work, we focus on the so-called ``single-photon ultrastrong-coupling regime'', where $g_{a,0}^{\omega}$ and $g_{d,0}^{\omega}$ are comparable in magnitude to the mechanical frequency $\Omega_{m}$. The Fano-mirror setup typically exhibits sufficiently strong optical coupling, such that \(\Lambda \gg |g_{a,0}^{\omega}|, |g_{d,0}^{\omega}|\)~\cite{WWoe2023}. This hierarchy allows us to treat the optomechanical interactions as perturbative corrections to the optical subsystem (we will verify this approximation numerically as shown in Fig.~\ref{Eliminate}). Accordingly, one can temporarily neglect the mechanical degrees of freedom and transform the Hamiltonian into the optical normal-mode basis, as in previous works focusing on the weak-coupling regime~\cite{Juliette2023pra,DL2025Fano}. Despite the large bare linewidths of the two optical modes, we will show that this setup can also \emph{effectively} enter the single-photon strong-coupling regime under appropriate conditions. It is important to note that the single-photon strong- and ultrastrong-coupling regimes are not mutually exclusive. The former refers to the situation where the single-photon optomechanical coupling strengths dominate over the optical dissipation rates, while the latter implies that these couplings approach or exceed the mechanical frequency. Therefore, a single optomechanical system can simultaneously lie in both regimes, which is the case for our effective model, as will be shown below.

Specifically, the optical Hamiltonian $\hat{H}_{\mathrm{opt}}$ can be diagonalized as
\begin{equation}
\hat{\tilde{H}}_\text{opt} = \omega_+ \hat{A}_+^\dagger \hat{A}_+ + \omega_- \hat{A}_-^\dagger \hat{A}_-,
\label{Hopttilde}
\end{equation}
where $\hat{A}_{\pm}$ denote the optical normal modes, defined as 
\begin{align}
\hat{A}_+ &= \cos \theta \; \hat{a} + \sin \theta \; \hat{d}, \nonumber\\
\hat{A}_- &= -\sin \theta \; \hat{a} + \cos \theta \; \hat{d}
\label{Aplusminus}
\end{align}
with the mixing angle $\theta$ determined by $\tan (2\theta) = \frac{2\Lambda}{\omega_a - \omega_d}$.
The corresponding eigenfrequencies of modes $A_{\pm}$ are given by
\begin{equation}
\omega_\pm = \frac{\omega_a + \omega_d}{2} \pm  \sqrt{\left(\frac{\omega_{a}-\omega_{d}}{2}\right)^2 + \Lambda^2}.
\label{omegapm}
\end{equation}

Moreover, by applying the inverse transformation
$\hat{a} = \cos{\theta} \hat{A}_+ - \sin{\theta} \hat{A}_-$ and $\hat{d} = \sin{\theta} \hat{A}_+ + \cos{\theta} \hat{A}_-$, the optomechanical interaction terms can be expressed in the normal-mode basis as
\begin{equation}
    \hat{\tilde{H}}_{\mathrm{om}} 
    = - \left[ G_+ \hat{A}_+^\dag \hat{A}_+ + G_- \hat{A}_-^\dag \hat{A}_- + G_{\text{mix}} \left(\hat{A}_+^\dag \hat{A}_- + \hat{A}_-^\dag \hat{A}_+ \right) \right]\left(\hat{b} + \hat{b}^{\dag}\right),
    \label{Homtilde}
\end{equation}
where the effective optomechanical coupling strengths are given by
\begin{align}
    G_+ &= g_{a,0}^{\omega} \cos^2\theta + g_{d,0}^{\omega} \sin^2\theta, \nonumber\\
    G_- &= g_{a,0}^{\omega} \sin^2\theta + g_{d,0}^{\omega} \cos^2\theta, \nonumber\\
    G_{\text{mix}} &= \frac{g_{d,0}^{\omega} - g_{a,0}^{\omega}}{2} \sin(2\theta).
    \label{Geffective}
\end{align}

Equation~(\ref{Geffective}) shows that the transformed Hamiltonian contains a mixed interaction term that simultaneously couples all three modes, with coupling strength $G_{\mathrm{mix}}$. Specifically, it describes a mechanically-induced tunneling (frequency conversion) process between the two normal modes. Such mixed interactions have been shown to enable intriguing physics in optomechanics~\cite{Harris2008nature}. For instance, in a membrane-in-the-middle optomechanical setup where the mixed optomechanical interaction dominates the system dynamics and when the normal-mode frequency splitting is tuned close to the mechanical frequency, both photon-phonon and photon-photon interactions can be significantly enhanced~\cite{LudwigPRL2012}. In the present work, however, we focus on regimes where only one of the optical normal modes contributes appreciably to the dynamics, and thereby the contribution of the mixed interaction is negligible. As we will show in Sec.~\ref{SecME}, under these conditions the transformed Hamiltonian can be greatly simplified.


\subsection{Effective master equation}\label{SecME}

Previous related works focus on the weak-coupling regime~\cite{Juliette2023pra,DL2025Fano} and use quantum Langevin equations for its description. By contrast, the system studied in this paper operates in the ultrastrong-coupling regime and, under appropriate conditions, can effectively access the strong-coupling regime. In this case, the standard linearization of the quantum Langevin equations becomes unreliable~\cite{Leijssen2017NC}, because the  quantum fluctuations of the relevant variables are no longer small compared to their mean values. More importantly, the resulting dynamics can become non-Gaussian in this regime, so that a Gaussian description based solely on first- and second-order moments is generally insufficient. We stress that this does not invalidate the full nonlinear quantum Langevin equations, which can provide a complete description of the system dynamics. However, they are less convenient for the nonlinear quantum effects considered below, such as photon blockade and mechanical cat-state generation, where access to the individual matrix elements of the density operator is often required. We therefore adopt a master-equation approach, which keeps the optomechanical interaction fully nonlinear and provides a direct framework for computing the density matrix and non-Gaussian observables. Several modified master equation formalisms have been developed for optomechanical systems in the single-photon strong- and ultrastrong-coupling regimes~\cite{HuDSME,NaseemGME}. The standard master-equation description, which assumes that the photonic (phononic) reservoir induces dissipation only in the optical (mechanical) degrees of freedom~\cite{RablPB2011,SVCR2}, remains accurate for quantum-optical observables studied in this paper, particularly when the mechanical damping rate $\gamma_m$ is small compared to the other relevant rates, such as the effective optical linewidth and optomechanical coupling strength; see Sec.~\ref{SecDSME} for details. In what follows, we adopt this standard description, which leads to the master equation
\begin{equation}
\dot{\hat{\rho}}=-i\left[\hat{H}_{\mathrm{tot}},\hat{\rho}\right]+\sum_{j=\{a,b,d,ad\}}\mathcal{L}_{j}[\hat{\rho}],
\label{MEexact}
\end{equation}
with the corresponding Lindblad dissipators defined as
\begin{subequations} \label{Lterms}
\begin{align}
\mathcal{L}_a[\hat{\rho}] &= \Gamma_a \left( \hat{a}\hat{\rho} \hat{a}^\dagger - \frac{1}{2}\left\{ \hat{a}^\dagger \hat{a}, \hat{\rho} \right\} \right), \label{La}\\
\mathcal{L}_d[\hat{\rho}] &= \kappa_d \left( \hat{d}\hat{\rho} \hat{d}^\dagger - \frac{1}{2}\left\{ \hat{d}^\dagger \hat{d}, \hat{\rho} \right\} \right), \label{Ld}\\
\mathcal{L}_{ad}[\hat{\rho}] &= \sqrt{\kappa_a\kappa_d} \left[ \hat{a}\hat{\rho}\hat{d}^\dagger - \frac{1}{2}\left\{ \hat{a}^\dagger\hat{d}, \hat{\rho} \right\} + \mathrm{H.c.} \right], \label{Lad}\\
\mathcal{L}_b[\hat{\rho}] &= \gamma_m(n_{\mathrm{th},b}+1) \left( \hat{b}\hat{\rho} \hat{b}^\dagger - \frac{1}{2}\left\{ \hat{b}^\dagger \hat{b}, \hat{\rho} \right\} \right)
+ \gamma_m n_{\mathrm{th},b} \left( \hat{b}^{\dag}\hat{\rho} \hat{b} - \frac{1}{2}\left\{ \hat{b} \hat{b}^{\dag}, \hat{\rho} \right\} \right). \label{Lb}
\end{align}
\end{subequations}
The dissipators in Eqs.~(\ref{La}), (\ref{Ld}), and (\ref{Lb}) describe the individual dissipation of modes $a$, $d$, and $b$, respectively, with $\Gamma_{a}=\kappa_{a}+\gamma_{a}$ the total decay rate of $a$. Equation~(\ref{Lb}) incorporates the effect of a finite thermal phonon population, where $n_{\mathrm{th},b}=[\mathrm{exp}(\hbar\Omega_{m}/(k_\mathrm{B}T))-1]^{-1}$ is the mean thermal phonon number with $k_\mathrm{B}$ the Boltzmann constant and $T$ the temperature. Thermal excitations of the optical modes can be safely neglected because their resonance frequencies are much higher than the energy scale set by temperature ($\hbar\omega_{a,d}\gg k_\mathrm{B} T$). The dissipation term in Eq.~(\ref{Lad}) accounts for the collective dissipation of modes $a$ and $d$, which effectively induces a \emph{dissipative interaction} $-i\frac{\sqrt{\kappa_a\kappa_d}}{2}\left(\hat{a}^\dag \hat{d} + \hat{d}^\dag \hat{a} \right)$. As mentioned above, this collective effect arises from the fact that both optical modes couple to a common photonic environment. It plays a crucial role in shaping the optical normal modes and is absent in conventional optomechanical systems. As a result, the overall (complex) coupling strength between $a$ and $d$ is effectively given by $\mathcal{G}=\Lambda-i\frac{\sqrt{\kappa_a\kappa_d}}{2}$~\cite{FanoM,WWoe2023,Juliette2023pra,DL2025Fano}. 

The dissipation part of the master equation \eqref{MEexact}, as given in Eqs.~\eqref{La}--\eqref{Lad}, is not in the canonical Lindblad form due to the term $\mathcal{L}_{ad}$. However, we show in Appendix~\ref{AppA:Lindblad} that it can be brought into the canonical Lindblad form and, therefore, gives rise to physically meaningful dynamics. Furthermore, the dynamics described by Eq.~\eqref{MEexact} is consistent with the dynamics described by the Langevin equations (before linearization) in Ref.~\cite{Juliette2023pra}, in the absence of dissipative optomechanical couplings and in the high-temperature limit, $k_\mathrm{B} T \gg \hbar \Omega_{m}$~\footnote{The optical part of the dissipation is exactly the same in both models, only the mechanical part is modeled slightly differently. In the master-equation approach, $\mathcal{L}_b$ [Eq.~\eqref{Lb}] corresponds to symmetric noise, while in the Langevin description the noise is often taken to act only on the momentum quadrature, see, e.g., Appendix~B.5 of Ref.~\cite{Juliette2021pra}}.

Similar to the procedure in Sec.~\ref{SecNormal}, the Lindblad dissipators in Eqs.~(\ref{La})--(\ref{Lad}) can also be transformed into the optical normal-mode basis. Due to the mixing of modes $\hat{a}$ and $\hat{d}$, the resulting expressions involve nontrivial cross-mode dissipative terms, including both individual decay channels for $\hat{A}_\pm$ and interference-like contributions between them. To provide a more intuitive understanding, we illustrate the system in the optical normal-mode representation in Fig.~\figpanel{Model}{c}, where all relevant dissipation and interaction channels are indicated. The full forms of the transformed dissipators are cumbersome, and for clarity, we present the explicit expressions in Appendix~\ref{AppA}. As will be shown below, these transformed dissipators can also be greatly simplified when the contribution of one optical normal mode is negligible.


We now consider a relevant regime, where only one of the optical normal modes contributes appreciably to the system dynamics, while the other mode, as well as the cross-mode terms, can be safely neglected. When the overall optical coupling $\mathcal{G}=\Lambda-i\sqrt{\kappa_a\kappa_d}/2$ is sufficiently strong, the two normal modes $\hat{A}_{\pm}$ may exhibit a large enough frequency splitting. In this case, if a coherent driving, which is near-resonant with one of the normal modes, is applied to the system (for instance, to mode $a$ via the right normal mirror), the excitation of the other normal mode becomes negligible due to the large detuning. Specifically, we consider a driving term $\hat{H}_{\mathrm{drive}}=\varepsilon_{p}\left(\hat{a}^\dag e^{-i\omega_{p}t} + \mathrm{H.c.}\right)$, where $\varepsilon_{p}$ and $\omega_{p}$ are the amplitude and frequency of the driving field, respectively. Transforming into the normal-mode basis and neglecting the contributions of $A_{+}$ (this approximation will be verified later), we obtain the effective Hamiltonian
\begin{equation}
\hat{H}_\text{eff} = \omega_- \hat{A}_-^\dagger \hat{A}_- + \Omega_m \hat{b}^{\dag}\hat{b} - G_{-} \hat{A}_-^\dagger \hat{A}_- \left(\hat{b} + \hat{b}^{\dag}\right) - \varepsilon_{p}\sin\theta\left(\hat{A}_{-}^{\dag}e^{-i\omega_p t} + \mathrm{H.c.}\right).
\label{twomodeH}
\end{equation}

Similarly, the effective Lindblad dissipator for mode $A_-$ can be obtained as (recall that the contributions of mode $A_{+}$ are neglected)
\begin{equation}\label{L-}
\mathcal{L}_{A_{-}}[\hat{\rho}] = \kappa_{-} \left[\hat{A}_{-}\hat{\rho}\hat{A}_{-}^{\dag} - \frac{1}{2}\left\{ \hat{A}_{-}^{\dag}\hat{A}_{-}, \hat{\rho} \right\} \right],
\end{equation}
with the effective decay rate given by
\begin{equation}
\kappa_{-} = \Gamma_a \sin^2 \theta + \kappa_d \cos^2 \theta - \sqrt{\kappa_a\kappa_d}\sin(2\theta).
\label{keff}
\end{equation}
Equation~(\ref{keff}) shows that, owing to the hybridization of the optical modes, $\kappa_{-}$ can be made much smaller than both $\Gamma_a$ and $\kappa_d$. This dissipation suppression leads to an effectively enhanced sideband resolution, which is beneficial for resolving optomechanical features at the single-photon level. Moreover, in the complementary case where the contribution of mode $A_-$ is negligible (when, e.g., the driving field is nearly resonant with mode $A_+$ instead), one can derive the effective decay rate of mode $A_{+}$ as
\begin{equation}
\kappa_{+} = \Gamma_a \cos^2 \theta + \kappa_d \sin^2 \theta + \sqrt{\kappa_a\kappa_d}\sin(2\theta),
\label{keff2}
\end{equation}
see Appendix~\ref{AppA} for details.
In this way, we arrive at a minimal optomechanical model that enables a range of genuine nonclassical phenomena, such as photon blockade and the generation of nonclassical mechanical states. 

\subsection{Relevant parameter regimes}\label{SecOpticalDynamics}

Before exploring nonclassical phenomena, we first identify parameter regimes in which the following two conditions are satisfied simultaneously: (\textit{i}) the excitation of one of the optical normal modes (e.g., $A_{+}$) can be neglected; and (\textit{ii}) the effective decay rate of the other normal mode (which dominates the system dynamics) is much smaller than the mechanical frequency $\Omega_{m}$. The first condition can be fulfilled when the driving field is resonant with the $A_{-}$ mode (i.e., $\omega_{p}=\omega_{-}$), and the frequency difference $\Delta_{\mathrm{normal}}=|\omega_{+}-\omega_{-}|$ between the two normal modes is much larger than the effective linewidth $\kappa_{+}$ of the $A_{+}$ mode. In this case, the $A_{+}$ mode remains far detuned from the driving field and is therefore barely excited. The second condition is more direct: it ensures that the optomechanical system operates in an effectively sideband-resolved regime~\cite{Juliette2023pra,DL2025Fano}, which is essential for observing a variety of nonclassical effects. 

To identify suitable parameter regimes, we numerically examine (using QuTiP~\cite{Qutip1,Qutip2}) the dynamics of the optical subsystem by solving the effective master equation
\begin{equation}
\dot{\hat{\rho}}_{\mathrm{opt}} \equiv \tilde{\mathcal{L}}_{\mathrm{opt}}[\hat{\rho}_{\mathrm{opt}}]
= -i\left[\hat{\tilde{H}}_{\mathrm{opt}}+\hat{\tilde{H}}_{\mathrm{drive}},\hat{\rho}_{\mathrm{opt}}\right] +\sum_{j=\{a,d,ad\}}\tilde{\mathcal{L}}_{j}[\hat{\rho}_{\mathrm{opt}}],
\label{optME}
\end{equation}
where $\hat{\rho}_{\mathrm{opt}}$ is the density matrix of the optical part, and $\hat{\tilde{H}}_{\mathrm{opt}}$, $\tilde{\mathcal{L}}_{a}$, $\tilde{\mathcal{L}}_{d}$, and $\tilde{\mathcal{L}}_{ad}$ are defined in Eqs.~(\ref{Hopttilde}), (\ref{Latilde}), (\ref{Ldtilde}), and (\ref{Ladtilde}), respectively. The effective driving Hamiltonian is given by
\begin{equation}
\hat{\tilde{H}}_{\mathrm{drive}}=\varepsilon_{p}\left[\left(\cos\theta\hat{A}_{+}^\dag-\sin\theta\hat{A}_{-}^{\dag}\right)e^{-i\omega_p t} + \mathrm{H.c.}\right],
\label{Hdrivetilde}
\end{equation}
which indicates that the $A_{+}$ and $A_{-}$ modes are driven with effective amplitudes $\varepsilon_{p,+}=\varepsilon_{p}\cos{\theta}$ and $\varepsilon_{p,-}=\varepsilon_{p}\sin{\theta}$, respectively. In Eqs.~(\ref{optME}) and (\ref{Hdrivetilde}), the tildes are used to distinguish quantities in the normal-mode basis from those in the original representation. Although $\varepsilon_{p,+}$ can be much larger than $\varepsilon_{p,-}$ for small $\theta$, the $A_{+}$ mode remains nearly unpopulated when conditions (\textit{i}) and (\textit{ii}) are simultaneously satisfied, because it is strongly detuned and subject to an extremely large linewidth. This will be verified numerically below.

\begin{figure}
\centering
\includegraphics[width = 0.55\linewidth]{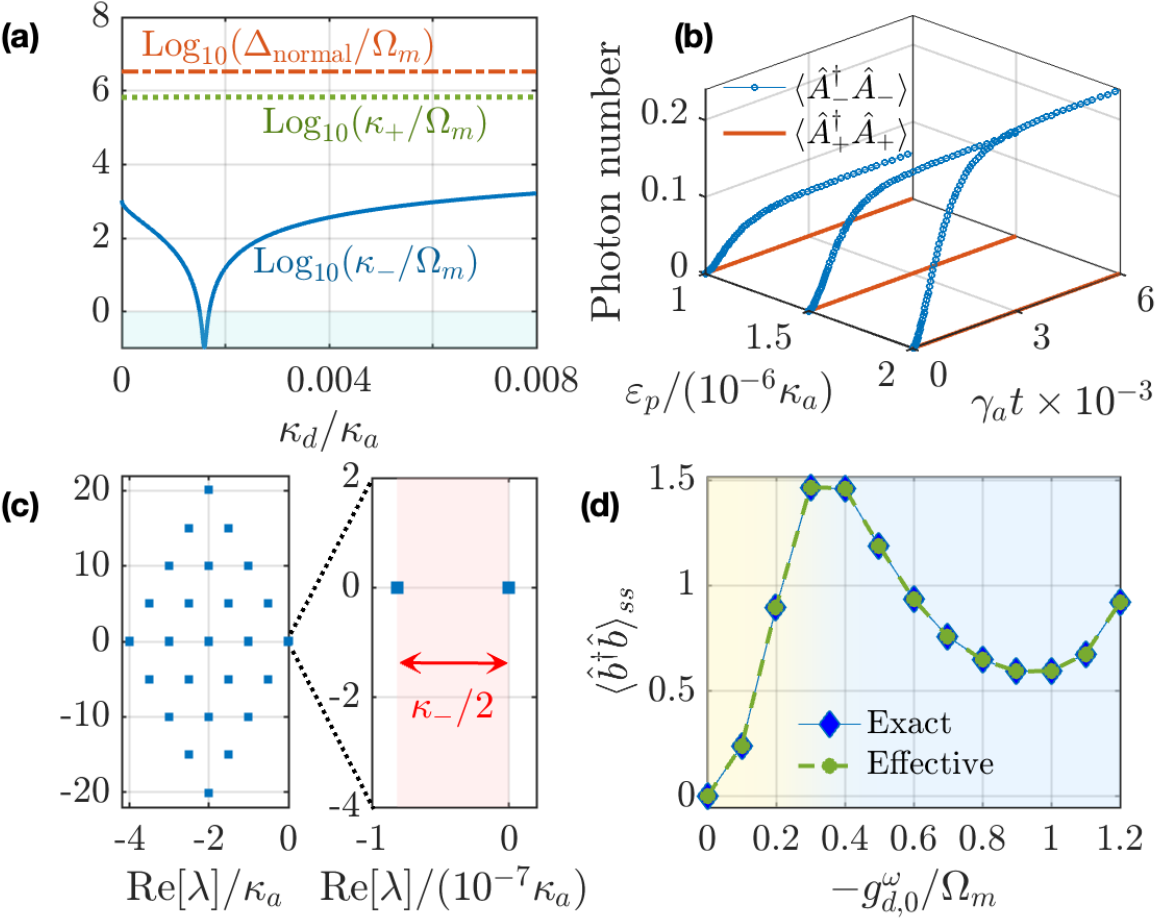}
\caption{(a) Effective parameters of the optical normal modes (on a logarithmic scale), derived from the effective master equation~(\ref{optME}), as functions of the dissipation ratio $\kappa_{d}/\kappa_{a}$. (b)~Time evolution of the mean photon numbers in the optical normal modes $A_{\pm}$, for $\kappa_{d}/\kappa_{a}=1.6\times10^{-3}$ and different values of the driving amplitude $\varepsilon_{p}$. (c) Eigenspectrum of the optical Liouvillian $\tilde{\mathcal{L}}_{\mathrm{opt}}$ for $\kappa_{d}/\kappa_{a}=1.6\times10^{-3}$. The left panel shows the full spectrum, while the right panel provides a zoom-in around the zero eigenvalue. The eigenvalues are denoted as $\lambda = \{\lambda_{1}, \lambda_{2}, \cdots\}$. (d) Steady-state mean phonon number $\langle b^{\dag}b \rangle_{ss}$ obtained from both the effective two-mode and the exact three-mode optomechanical models, as a function of $-g_{d,0}^{\omega}$ (normalized by the mechanical frequency), with $\varepsilon_{p}/\kappa_a=1\times10^{-8}$ and $\kappa_{d}/\kappa_{a}=1.6\times10^{-3}$. In both panels (b) and (d), the driving field is chosen to be resonant with the $A_-$ mode, i.e., $\omega_{p}=\omega_{-}$. Other parameters: $\gamma_a/\kappa_a = 1\times10^{-4}$, $\omega_a/\kappa_a = 200$, $\omega_d/\kappa_a = 195$, $\Lambda/\kappa_a = 0.2$, $\Omega_{m}/\kappa_a=1.5\times10^{-6}$, $\gamma_{m}/\kappa_a=8\times10^{-14}$, $g_{a,0}^{\omega}/\Omega_{m}=0.25$, and $n_{\mathrm{th},b}=0$.}
\label{Eliminate}
\end{figure}

In Fig.~\figpanel{Eliminate}{a}, we plot the effective parameters in the optical normal-mode basis, including the frequency difference $\Delta_{\mathrm{normal}}$ and the effective decay rates $\kappa_{\pm}$, as derived from the master equation~(\ref{optME}). The chosen parameters are experimentally accessible and similar to those considered in Ref.~\cite{Juliette2023pra}. Using this parameter set, both conditions (\textit{i}) and (\textit{ii}) can be satisfied. Note that the mixing angle is small ($\theta\approx0.04$) due to the large frequency difference $|\omega_{a}-\omega_{d}|$ compared to the optical coherent coupling $\Lambda$. As a result, in the effective Hamiltonian~(\ref{twomodeH}) the effective optomechanical coupling strength $G_{-}$ is essentially set by $g_{d,0}^{\omega}$ for the parameters considered. Quantitatively, their difference $|g_{d,0}^{\omega}-G_{-}|$ remains at the order of $10^{-3}$ (in units of $\Omega_m$) across the parameter range of interest. We further emphasize that the parameter set considered in Fig.~\figpanel{Eliminate}{a} is not unique for our purpose; an alternative viable choice is presented in Appendix~\ref{AppB}. Additional parameter options have also been explored in Refs.~\cite{Juliette2023pra,DL2025Fano}.

Note that the effective parameters of the optical normal modes can be alternatively determined within a full (non-linearized) Langevin-equation framework~\cite{Juliette2023pra,DL2025Fano}. In that framework, the environmental degrees of freedom are eliminated and the system dynamics are governed by the resulting quantum Langevin equations. In Appendix~\ref{AppB}, we provide more details on the Langevin-equation framework and compare the effective parameters extracted from the master equation (\ref{optME}) with those obtained from the Langevin-equation framework. The two sets of results show excellent agreement within the parameter regime of interest, supporting the validity of the effective master equation formalism adopted in this work.

To validate the theoretical prediction, in Fig.~\figpanel{Eliminate}{b} we show the time evolution of the mean photon numbers in the two optical normal modes, with different values of $\varepsilon_{p}$ and all other parameters identical to Fig.~\figpanel{Eliminate}{a}. Applying a weak driving field that is resonant with the $A_{-}$ mode, we observe that the $A_{+}$ mode remains nearly unexcited throughout the entire evolution. With the parameters considered, the $A_{+}$ mode reaches steady-state populations only at the level of $10^{-11}$ for all three drive amplitudes shown. In contrast, the $A_{-}$ mode is gradually populated and evolves toward a steady state with a finite mean photon number. This steady-state photon number can be tuned over a reasonable range by varying the driving amplitude $\varepsilon_{p}$, as shown by the three representative curves (solid lines with circle markers) corresponding to different $\varepsilon_{p}$. Although the bare driving amplitude appears much larger than the effective decay rate $\kappa_{-}$ (by more than an order of magnitude), the \emph{effective} driving amplitude experienced by the $A_-$ mode is actually $\varepsilon_{p,-}=\varepsilon_{p}\sin\theta\sim\kappa_{-}$. As a result, the steady state of $A_-$ approximates a weak coherent state with a small field amplitude. To further support our analysis, we plot the eigenspectrum of the Liouvillian $\tilde{\mathcal{L}}_{\mathrm{opt}}$ [cf. Eq.~(\ref{optME})] in Fig.~\figpanel{Eliminate}{c}, where the eigenvalues are denoted as $\lambda = \{\lambda_{1}, \lambda_{2}, \cdots\}$. All eigenvalues, except for the zero eigenvalue associated with the steady state, have negative real parts. This confirms that the system asymptotically approaches a unique nontrivial steady state. The spectral gap, which is defined as the smallest nonzero value of 
\(-{\rm Re}[\lambda]\), is found to coincide with 
\(\kappa_-/2\) ($\kappa_{-}/\kappa_{a}\approx1.633\times10^{-7}$ for the parameters considered), see the inset in Fig.~\figpanel{Eliminate}{c}. This agreement further verifies that the steady-state dynamics are predominantly governed by the $A_-$ mode.

Before proceeding, we incorporate the mechanical mode and the associated optomechanical interactions into our analysis, and assess the validity of reducing the full system to an effective two-mode optomechanical model. This verification is performed by comparing the steady-state mean phonon number, $\langle b^{\dag}b \rangle_{\mathrm{ss}}$, obtained from both the effective two-mode master equation
\begin{equation}
\dot{\hat{\rho}} = -i[\hat{H}_{\mathrm{eff}},\hat{\rho}] + \sum_{j=A_{-},b} \mathcal{L}_{j}[\hat{\rho}],
\label{MEeff}
\end{equation}
and the exact three-mode master equation~(\ref{MEexact}), with the corresponding driving term included. As shown in Fig.~\figpanel{Eliminate}{d}, the two approaches yield excellent agreement across a wide range of the normalized optomechanical coupling strength $-g_{d,0}^{\omega}/\Omega_{m}$ ($g_{d,0}^{\omega}$ typically takes negative values in experiments~\cite{WWoe2023}). This confirms the validity of the mode-elimination approximation under appropriate conditions. The non-monotonic dependence of the steady-state phonon number on $|g_{d,0}^{\omega}|$ can be understood from the fact that, increasing the optomechanical coupling not only enhances the radiation-pressure backaction but also reshapes the optical steady state that drives the mechanics. In particular, a turning point can emerge [at the crossing of the yellow- and blue-shaded regions in Fig.~\figpanel{Eliminate}{d}] when the intrinsic single-photon nonlinearity becomes resolvable (when $G_-^{2}/\Omega_m > \kappa_{-}$, i.e., $-g_{d,0}^{\omega}/\Omega_{m}\gtrsim0.3$, as explained in Sec.~\ref{SecPB}). This marks a crossover from an essentially linear response to a regime where nonlinear features strongly affect the mechanical state.

\section{Results}

\subsection{Photon blockade effect}\label{SecPB}

\begin{figure}
\centering
\includegraphics[width = 0.6\linewidth]{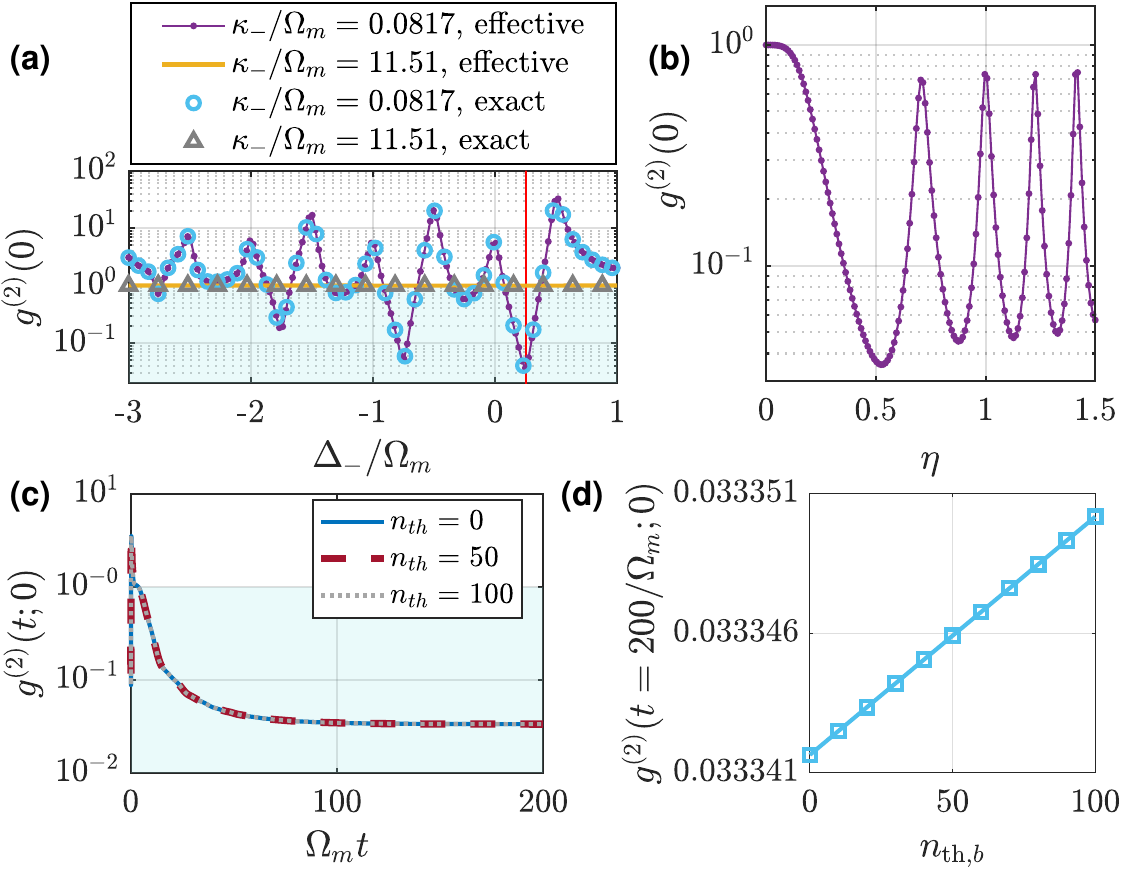}
\caption{Photon blockade in the dark normal mode. (a) Steady-state value $g^{(2)}(0)$ of the equal-time second-order correlation function versus the detuning $\Delta_{-}$ (between the $A_-$ mode and the driving field) for two values of $\kappa_{-}$. With the parameters used here, $\kappa_{-}/\Omega_m=0.0817$ and $11.51$ correspond to $\kappa_d/\kappa_a=1.6\times10^{-3}$ and $2.0\times10^{-3}$, respectively; see also Fig.~\figpanel{Eliminate}{a}. For comparison, we provide the results obtained from both the two-mode effective model (described by $\hat{H}_{\mathrm{eff}}$) and the three-mode exact model (described by $\hat{H}_{\mathrm{tot}}+\hat{H}_{\mathrm{drive}}$). The red vertical line marks $\Delta_{-}=G_{-}^{2}/\Omega_{m}$, where $g^{(2)}(0)$ reaches its minimum. (b) $g^{(2)}(0)$ as a function of the normalized coupling strength $\eta=-G_{-}/\Omega_{m}$. (c) Time evolution of the equal-time second-order correlation function $g^{(2)}(t;0)$ for different mean thermal phonon numbers $n_{\mathrm{th},b}$. (d) $g^{(2)}(t;0)$ evaluated at $\Omega_{m}t=200$ as a function of $n_{\mathrm{th},b}$. The light blue areas in (a) and (c) represent the region where the photon antibunching effect arises. Parameters: $n_{\mathrm{th},b}=0$ in (a) and (b); $\eta=0.5$ in (a), (c), and (d); $\kappa_{d}/\kappa_{a}=1.6\times10^{-3}$ and $\Delta_{-}=G_{-}^{2}/\Omega_{m}$ in (b)--(d). Other parameters are $\gamma_a/\kappa_a = 1\times10^{-4}$, $\omega_a/\kappa_a = 200$, $\omega_d/\kappa_a = 195$, $\Lambda/\kappa_a = 0.2$, $\Omega_{m}/\kappa_a = 2\times10^{-6}$, $\gamma_{m}/\kappa_a = 8\times10^{-14}$, $\varepsilon_{p}/\kappa_a = 2\times10^{-9}$, and $g_{a,0}^{\omega}/\Omega_{m}=0.25$. }
\label{Blockade}
\end{figure}

To investigate the performance of the Fano-mirror optomechanical setup in the single-photon (ultra)strong coupling regime, we begin by examining the photon statistics, under the condition that the $A_+$ mode is effectively eliminated, as identified in Sec.~\ref{SecOpticalDynamics}. In particular, we focus on the photon blockade effect~\cite{RablPB2011} in the $A_-$ mode, which manifests as a suppression of multiphoton occupation in this normal mode. This effect is expected to occur when the optomechanically induced anharmonicity becomes spectrally resolvable, i.e., when $G_{-}^{2}/\Omega_{m}>\kappa_{-}$ (as explained below). This condition ensures that the effective model described by $\hat{H}_{\mathrm{eff}}$ operates in the sideband-resolved, single-photon strong-coupling regime, where photon blockade is experimentally observable.  

To test this prediction, we numerically solve the master equation~(\ref{MEeff}) and evaluate the equal-time second-order correlation function~\cite{RablPB2011}
\begin{equation}
g^{(2)}(t;0)= \frac{\left\langle \hat{A}_{-}^{\dag}(t)\hat{A}_{-}^{\dag}(t)\hat{A}_{-}(t)\hat{A}_{-}(t) \right\rangle}{\left\langle \hat{A}_{-}^{\dag}(t)\hat{A}_{-}(t) \right\rangle^{2}}.
\label{g20}
\end{equation}
We denote its steady-state value as $g^{(2)}(0) \equiv \mathrm{lim}_{t\rightarrow\infty}\, g^{(2)}(t;0)$. This quantity provides a direct experimental probe of photon statistics.  While a coherent state exhibits Poissonian statistics with $g^{(2)}(0)=1$, photon bunching and antibunching are indicated by $g^{(2)}(0)>1$ and $g^{(2)}(0)<1$, respectively. In particular, photon blockade occurs when $g^{(2)}(0) \rightarrow 0$.

In Fig.~\ref{Blockade}, we demonstrate the photon-blockade effect in the $A_{-}$ mode. As shown in Fig.~\figpanel{Blockade}{a}, when $\kappa_{-}\ll\Omega_{m}$ (purple dot-solid line), a series of bunching and antibunching resonances appears, with the strongest antibunching observed at $\Delta_{-}=\omega_{-}-\omega_{p}=G_{-}^{2}/\Omega_{m}$. By contrast, these resonances disappear when the sideband resolution is lost, i.e., when $\kappa_{-}/\Omega_{m}\gg 1$ (orange dashed line). We again benchmark the effective two-mode model against the exact three-mode description, as shown by the excellent agreement between the curves and markers. This agreement further confirms the validity of the effective two-mode Hamiltonian in Eq.~(\ref{twomodeH}) within the considered parameter regime. The result in Fig.~\figpanel{Blockade}{a} can be understood via the polaron transform 
\begin{equation}
\hat{U}^{\dag}\hat{H}_{\mathrm{eff}}\hat{U} = \left(\omega_{-}-\frac{G_{-}^{2}}{\Omega_{m}}\right) \hat{A}^{\dag}_{-}\hat{A}_{-} + \Omega_{m}\hat{b}^{\dag}\hat{b} - \frac{G_{-}^{2}}{\Omega_{m}} \hat{A}_{-}^{\dag}\hat{A}_{-}^{\dag}\hat{A}_{-}\hat{A}_{-} - \varepsilon_{p}\sin{\theta}\left[\hat{D}\left(\frac{G_{-}}{\Omega_{m}}\right)\hat{A}_{-}e^{i\omega_{p}t} + \text{H.c.}\right] 
\label{UHeff}
\end{equation}
with $\hat{U}=\mathrm{exp}\left[G_{-}\hat{A}_{-}^{\dag}\hat{A}_{-}\left(\hat{b}^{\dag}-\hat{b}\right)/\Omega_{m}\right]$, and the displacement operator $\hat{D}(\alpha)=\mathrm{exp}\left(\alpha \hat{b}^{\dag} - \alpha^{*}\hat{b}\right)$. The transformed Hamiltonian reveals that the Fock states $|n_{-}\rangle$ of the $A_-$ mode acquire an energy shift of $-(n_{-}G_{-})^{2}/\Omega_{m}$, rendering this mode effectively anharmonic with a Kerr-type nonlinearity. Consequently, when the driving field is resonant with the $|0_{-}\rangle\leftrightarrow|1_{-}\rangle$ transition, i.e., $\Delta_{-}=G_{-}^{2}/\Omega_{m}$, it becomes detuned from the subsequent transition $|1_{-}\rangle\leftrightarrow|2_{-}\rangle$, which is thereby suppressed if this single-photon nonlinearity is resolvable, i.e., $G_{-}^{2}/\Omega_{m} > \kappa_{-}$. Note that different phonon sidebands are no longer spectrally separated in the sideband-unresolved regime ($\Omega_m \lesssim \kappa_{-}$). This washes out the bunching/antibunching resonance structure and thus suppresses the observable photon blockade.


We next define a dimensionless optomechanical coupling strength $\eta=-G_{-}/\Omega_{m}$ (recall that $G_{-}\approx g_{d,0}^{\omega}<0$ in experiments) and investigate how the effective optomechanical interaction influences the photon blockade effect. In Fig.~\figpanel{Blockade}{b}, we plot $g^{(2)}(0)$ as a function of $\eta$. It shows that $g^{(2)}(0)$ reaches a minimum near $\eta=0.5$ and exhibits a sequence of pronounced peaks as $\eta$ increases further. This indicates that stronger optomechanical coupling does not necessarily enhance the photon blockade effect~\cite{RablPB2011}. Beyond the optimal point, increasing $\eta$ can instead enhance the multiphoton resonances and degrade photon blockade. 

We further examine the sensitivity of photon blockade to thermal phonon excitations, which are a common experimental limitation in optomechanics. In Figs.~\figpanel{Blockade}{c} and \figpanel{Blockade}{d}, we show the time evolution of $g^{(2)}(t;0)$ and its value at $\Omega_{m}t=200$, respectively, for different mean thermal phonon numbers $n_{\mathrm{th},b}$, assuming that the mechanical mode is initially precooled. We find that $g^{(2)}(t;0)$ approaches a nearly steady value after $\Omega_{m}t\approx100$, and that the dynamics is quite insensitive to $n_{\mathrm{th},b}$ over the shown timescale. This robustness can be understood from the sufficiently small mechanical damping rate $\gamma_{m}$ (here $\gamma_{m}/\kappa_{-}\approx4.9\times10^{-7}$), which sets the timescale for mechanical dissipation and (re)thermalization. Specifically, for an initially precooled mechanical mode the thermalization rate is approximately given by $\gamma_{m}n_{\mathrm{th},b}$. In the parameter regime considered here, this implies that the thermalization effect can be neglected during the evolution shown. This robustness diminishes gradually as $\gamma_m$ increases~\cite{LiaoUSC2020,YinUSC2022}.


In experiments, the photon statistics of the $A_-$ mode can be inferred from photon counting and coincidence measurements performed on the output fields emerging from the left PhC mirror~\cite{RablPB2011}. Importantly, because the PhC mirror hosts the Fano mode, the left output field generally contains contributions from both optical modes $a$ and $d$, i.e., $\hat{o}_{\mathrm{out}}=\sqrt{\kappa_{a}}\hat{a} + \sqrt{\kappa_{d}}\hat{d} -\hat{o}_{\mathrm{in}}$ with $\hat{o}_{\mathrm{in}}$ the left input field operator~\cite{Juliette2023pra,DL2025Fano}. In the ideal situation where only the $A_{-}$ mode is populated while the $A_{+}$ mode remains in the vacuum state, the two original optical modes can be expressed as
$\hat{a} = -\sin\theta \, \hat{A}_{-}$ and
$\hat{d} = \cos\theta \, \hat{A}_{-}$, such that $\sqrt{\kappa_{a}}\hat{a} + \sqrt{\kappa_{d}}\hat{d} \propto \hat{A}_{-}$.
In our protocol, the left PhC mirror is assumed to be driven only by vacuum fluctuations. Under this condition, all normally ordered moments involving $\hat{o}_{\mathrm{in}}$ do not contribute, and the output-field correlation functions [obtained by replacing $\hat{A}_{-}$ with $\hat{o}_{\mathrm{out}}$ in Eq.~\eqref{g20}] are entirely determined by the intracavity operators. According to the definition in Eq.~(\ref{g20}), any proportionality factor $\alpha$ relating the output operators to the intracavity ones (e.g., $\hat{o}_{\mathrm{out}} = \alpha \hat{o}$) cancels out between the numerator and denominator. Therefore, in the ideal case, the second-order correlation function extracted from the left output field coincides with that of the $A_{-}$ mode. Deviations from this equivalence may arise if the $A_{+}$ mode becomes non-negligibly populated. While Fig.~\ref{Blockade} focuses on the weak-driving limit, in an experiment one may deliberately increase the drive amplitude to boost the detected photon flux and thereby shorten the integration time. This comes at the price of a larger $g^{(2)}(0)$, but there exists an optimal operating point that optimizes the trade-off between the antibunching contrast $1-g^{(2)}(0)$ and the detected photon flux, enabling the fastest measurement. For the parameters of Fig.~\ref{Blockade}, we have verified that increasing the drive strength by a factor of $200$ raises $g^{(2)}(0)$ to $\approx 0.6$ at $\eta=0.5$, while the output photon flux grows by roughly two orders of magnitude. Operating in this brighter regime still yields clear antibunching and enables substantially faster data acquisition.

\subsection{Non-Gaussian mechanical states}\label{SecNonGaussian}

In Sec.~\ref{SecPB}, we investigated the nonclassical photon statistics enabled by the effective sideband-resolved and single-photon strong-coupling characteristics in our system. In this section, we demonstrate the feasibility of preparing non-Gaussian mechanical states within the same framework. In the absence of external driving fields, Eq.~(\ref{UHeff}) describes an optical mode with Kerr-type nonlinearity and a (decoupled) linear mechanical harmonic oscillator. In this case, any initially Gaussian mechanical state remains Gaussian throughout the evolution. To generate non-Gaussian mechanical states, appropriate driving protocols are generally required, as they can introduce effective anharmonicity to the mechanical mode. In view of this, we adopt the \emph{bichromatic driving} scheme proposed in Ref.~\cite{Hauer2023prl}, which enables high-fidelity generation of motional cat states in the single-photon strong-coupling regime. Specifically, we apply two coherent driving fields: one resonant with the $A_{-}$ mode (i.e., $\omega_{p,1} = \omega_{-}$) and the other red-detuned from $A_{-}$ by approximately twice the mechanical frequency (i.e., $\Delta_{-}' = \omega_{p,2} - \omega_{-} \approx -2\Omega_{m}$). The detuned drive induces an effective \emph{two-phonon cooling process}, which is a key ingredient for mechanical cat-state generation~\cite{Hauer2023prl}. 

With the bichromatic driving scheme, the Hamiltonian of the effective model reads
\begin{equation}
\hat{H}_{\mathrm{bi}} = \omega_{-} \hat{A}_{-}^{\dag}\hat{A}_{-} + \Omega_{m}\hat{b}^{\dag}\hat{b} - G_{-}\hat{A}_{-}^{\dag}\hat{A}_{-}\left(\hat{b}+\hat{b}^{\dag}\right) - \sum_{l=1,2}\varepsilon_{p,l}\sin{\theta} \left(\hat{A}_{-}^{\dag}e^{-i\omega_{p,l}t} + \hat{A}_{-}e^{i\omega_{p,l}t}\right),
\label{Htwotone1}
\end{equation}
where $\varepsilon_{p,1}$ and $\varepsilon_{p,2}$ denote the amplitudes of the resonant and detuned driving fields, respectively. For numerical simulations, we displace the $A_-$ mode by its steady-state amplitude, which is determined by the driving amplitude $\varepsilon_{p,2}$, i.e., $\alpha_{-} = \varepsilon_{p,2}\sin{\theta}/(\Delta_{-}'+i\kappa_{-}/2)$. This displacement, along with an extra frame rotation at $\omega_{p,2}$, is implemented via the unitary transformation $\hat{U}'=\text{exp}(i\omega_{p,2}\hat{A}_{-}^{\dag}\hat{A}_{-}t)\text{exp}(\alpha_{-}\hat{A}_{-}^{\dag}-\alpha_{-}^{*}\hat{A}_{-})$. In this transformed frame, the Hamiltonian becomes
\begin{eqnarray}
\hat{H}_{\mathrm{bi}}' &=& -\Delta_{-}'\hat{A}_{-}^{\dag}\hat{A}_{-} + \Omega_{m}\hat{b}^{\dag}\hat{b} - G_{-}\hat{A}_{-}^{\dag}\hat{A}_{-}\left(\hat{b}+\hat{b}^{\dag}\right) - G_{-}\left(\alpha_{-}\hat{A}_{-}^{\dag} + \alpha_{-}^{*}\hat{A}_{-}\right)\left(\hat{b} + \hat{b}^{\dag}\right) \nonumber\\
&&- \varepsilon_{p,1}\sin{\theta} \left(\hat{A}_{-}^{\dag}e^{i\Delta_{2,1}t} + \hat{A}_{-}e^{-i\Delta_{2,1}t}\right),
\label{Htwotone2}
\end{eqnarray}
where $\Delta_{2,1}=\omega_{p,2}-\omega_{p,1}$ is the detuning between the two driving fields. Note that in Eq.~(\ref{Htwotone2}), we have omitted the term $-G_{-}|\alpha_{-}|^{2} \left(\hat{b}+\hat{b}^{\dag}\right)$ by applying an additional displacement transformation $\hat{D}(\beta)=\text{exp}\left(\beta \hat{b}^{\dag} - \beta^{*} \hat{b}\right)$, with $\beta=G_{-}|\alpha_{-}|^{2}/\Omega_{m}$. The induced static frequency shift of the $A_-$ mode, $\delta_{-}=-2G_{-}^{2}|\alpha_{-}|^{2}/\Omega_{m}$, is then absorbed into the definition of the detuned driving frequency. With these transformations, one can numerically solve the system dynamics with the master equation 
\begin{equation}
\dot{\hat{\rho}} = -i\left[\hat{H}_{\mathrm{bi}}', \hat{\rho}\right] + \sum_{j=A_{-},b} \mathcal{L}_{j}[\hat{\rho}].
\label{MEcat}
\end{equation}

\begin{figure*}
\centering
\includegraphics[width = 0.94 \linewidth]{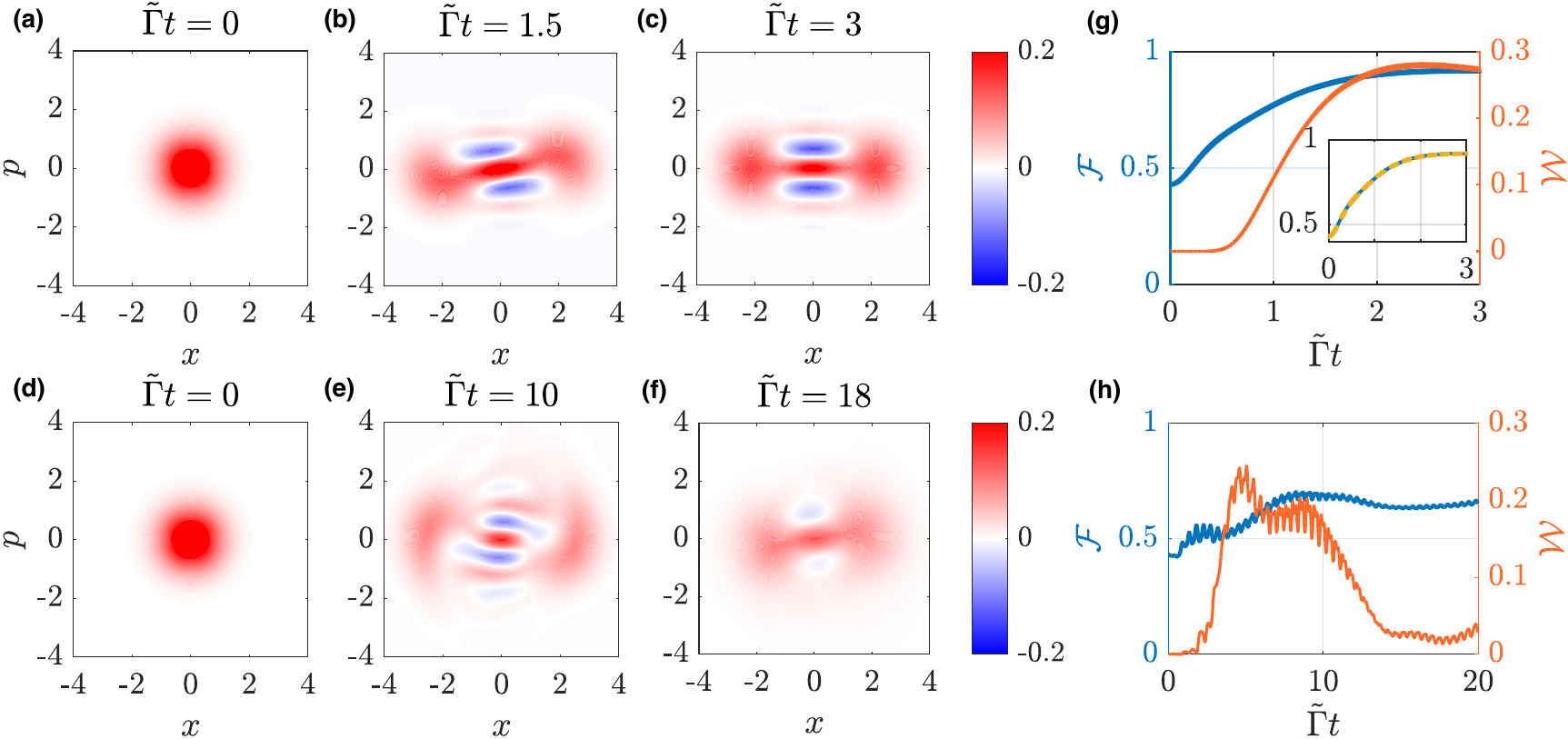}
\caption{Preparation processes of non-Gaussian mechanical states. (a)--(f) Snapshots of the Wigner function $W(x,p)$ of the mechanical mode at different times (normalized by $\tilde{\Gamma}=4|\tilde{G}_{-}|^{2}/\kappa_{-}$, with $\tilde G_-=G_-^2\alpha_-/\Omega_m$), with [(a)--(c)] $\eta=0.15$ and [(d)--(f)] $\eta=0.35$. (g) and (h) Fidelity $\mathcal{F}$ and Wigner logarithmic negativity $\mathcal{W}$ of the prepared mechanical state with respect to the ideal even cat state [defined as $|\mathrm{cat},+\rangle$ in Eq.~(\ref{catstates})], with (g) $\eta=0.15$ and (h) $\eta=0.35$. The inset in (g) compares the time evolution of the fidelity $\mathcal{F}$ with $n_{\mathrm{th},b}=0$ (blue solid) and $n_{\mathrm{th},b}=100$ (yellow dashed), with all other parameters identical to the main panel. In both cases (i.e., $\eta=0.15$ and $\eta=0.35$), the mechanical mode is initially prepared in the vacuum state. Other parameters (unless otherwise specified) are $\gamma_a/\kappa_a = 1\times10^{-4}$, $\omega_a/\kappa_a = 200$, $\omega_d/\kappa_a = 195$, $\Lambda/\kappa_a = 0.2$, $\Omega_{m}/\kappa_a = 4\times10^{-6}$, $\gamma_{m}/\kappa_a = 8\times10^{-14}$, $g_{a,0}^{\omega}/\Omega_{m}=0.12$, $\alpha_{-}=\sqrt{0.1}$, $n_{\mathrm{th},b}=0$, and $\chi=1.54$.}
\label{CatS}
\end{figure*}

According to Ref.~\cite{Hauer2023prl}, a two-component (two-legged) mechanical cat state can be generated when the optomechanical system operates in the single-photon strong-coupling regime, i.e., when $|G_{-}|>\kappa_{-}/2$ in our setup, which is clearly satisfied as demonstrated above. To prepare the cat states 
\begin{equation}
|\mathrm{cat},\pm\rangle=\frac{\left(|\chi\rangle \pm |-\chi\rangle\right)}{\sqrt{2\left(1\pm e^{-2|\chi|^{2}}\right)}},
\label{catstates}
\end{equation}
where $|\chi\rangle$ denotes a coherent state of the mechanical mode with amplitude $\chi$, we apply a resonant driving field with $\varepsilon_{p,1}=G_{-}^{2}\chi^{2}\alpha_{-}/\Omega_{m}$~\cite{Hauer2023prl}.
Figure~\ref{CatS} shows snapshots of the Wigner function~\cite{Wigner1932} of the mechanical state, which is defined as
\begin{equation}
W(x, p) = \frac{1}{\pi} \int_{-\infty}^{\infty} \langle x + z | \hat{\rho}_{b} | x - z \rangle \, e^{-2ipz} \, dz
\label{Wmec}
\end{equation}
with $\hat{\rho}_{b}=\mathrm{Tr}_{A_-}\left(\hat{\rho}\right)$ being the reduced density matrix of the mechanical mode. Indeed, for smaller $\eta$, the mechanical mode evolves from the vacuum towards an even cat state [cf. Figs.~\figpanel{CatS}{a}--\figpanel{CatS}{c}]. However, as $\eta$ becomes larger, the prepared state departs from an ideal cat state [cf. Figs.~\figpanel{CatS}{d}--\figpanel{CatS}{f}], even though the condition $|G_{-}|>\kappa_{-}/2$ still holds.

This deviation can be understood by performing the unitary transformation $\hat{U}$ defined in Eq.~(\ref{UHeff}) on the transformed optomechanical coupling term $G_{-}\hat{A}_{-}^{\dag}\hat{A}_{-}\left(\hat{b}+\hat{b}^{\dag}\right)$ in Eq.~\eqref{Htwotone2}, which yields $\left(\tilde{G}_{-}^{*}\hat{A}_{-} - \tilde{G}_{-}\hat{A}_{-}^{\dag}\right)\left(\hat{b}^2 - \hat{b}^{\dag 2}\right) + \mathcal{O}(G_{-}^{2}/\Omega_{m})$ to leading order in $G_{-}/\Omega_{m}$~\cite{Hauer2023prl}. It shows that the nonlinear optomechanical interaction results in a second-order phonon process with an effective rate $\tilde{G}_{-}=G_{-}^{2}\alpha_{-}/\Omega_{m}$. The remaining term $\mathcal{O}(G_{-}^{2}/\Omega_{m})$ describes the higher-order nonlinear corrections, which are proportional to $G_{-}^{2}/\Omega_{m}$. However, the above analysis relies on the assumption $|G_{-}|/\Omega_{m} \ll 1$ (i.e., $\eta \ll 1$ under our current parameters), such that $G_{-}/\Omega_{m}$ can be treated as a small perturbative parameter and only the leading-order terms are retained. When $|G_{-}|$ becomes comparable to $\Omega_{m}$, this approximation breaks down: the effective two-phonon description (enabling the cooling and stabilization of the cat state) becomes inaccurate, and higher-order nonlinearities (e.g., three- or four-phonon processes) can no longer be neglected. In this regime, the system dynamics becomes too complex to maintain a cat state.

To quantitatively evaluate the prepared mechanical state, we calculate the fidelity $\mathcal{F}=\sqrt{\langle\mathrm{cat}, +|\hat{\rho}_{b}|\mathrm{cat},+\rangle}$ of the mechanical state $\hat{\rho}_{b}$ with respect to the ideal even cat state $\hat{\rho}_{\mathrm{cat},+}=|\mathrm{cat},+\rangle \langle\mathrm{cat},+|$. We also quantify the nonclassicality of the prepared mechanical state by calculating the Wigner logarithmic negativity~\cite{WLN2018}
$\mathcal{W}=\ln\!\left[\iint dxdp |W(x,p)|\right]$,
which vanishes for states with a non-negative Wigner function (in particular, all Gaussian states) and takes positive values for states exhibiting Wigner negativity. Larger values of $\mathcal{W}$ indicate stronger Wigner-function negativity, and thus a higher degree of nonclassicality. 
Note that the parity of the prepared cat state is determined by that of the initial mechanical state. If the mechanical mode is initialized in the odd Fock-state manifold, one would instead expect the generation of an odd cat state. The dynamics of the fidelity and the Wigner logarithmic negativity for two representative values of $\eta$ are shown in Figs.~\figpanel{CatS}{g} and \figpanel{CatS}{h}. For a moderate $\eta$, the protocol yields a high-fidelity cat state together with a considerable degree of nonclassicality, as expected. For larger $\eta$, however, the prepared mechanical state deviates significantly from a standard cat state. In particular, $\mathcal{W}$ exhibits a relatively large transient peak but then drops drastically, indicating that the nonclassicality is not stably maintained. We also compare the dynamics of $\mathcal{F}$ for $\eta=0.15$ in the cases $n_{\mathrm{th},b}=0$ and $n_{\mathrm{th},b}=100$, as shown in the inset of Fig.~\figpanel{CatS}{g}. The two traces are nearly indistinguishable, confirming that mechanical thermalization is negligible on the state-preparation timescale owing to the ultralow mechanical damping rate.

After preparing the mechanical cat state, it can be read out using established mechanical-state tomography protocols that remain compatible with our effective model. One direct route is quadrature tomography based on back-action–evading measurements~\cite{VannerTomography2014}. Such quadrature measurements have been implemented in resolved-sideband platforms and used for mechanical-state tomography~\cite{ShomroniTomography2019}. As a complementary strategy that is well suited to the single-photon (ultra)strong-coupling regime, one may exploit single-photon scattering/emission spectroscopy~\cite{LiaoTomography2014}. In that regime, the spectrum of scattered or emitted photons encodes information about the underlying mechanical state over a wide range of cavity-field decay rates and does not rely on the sideband-resolved condition. This strategy provides a route to reconstruct (or strongly constrain) the mechanical phonon-state statistics and to benchmark nonclassicality even when operating deep in the nonlinear regime.

We note that alternative measures of cat-state quality can also be employed. Recently, ``catability'' was introduced as a directly measurable criterion based on nonlinear squeezing, which quantifies cat-like non-Gaussian features by comparing the measured state against the optimal Gaussian benchmark~\cite{Catability2026}. This provides an appealing alternative to fidelity, particularly in experimental situations where full quantum-state tomography is demanding. More details on catability and representative numerical simulations are provided in Appendix~\ref{AppC}.


\section{Discussion}

\subsection{Impact of dissipative optomechanical couplings}\label{SecDissipative}

So far, we have considered an idealized case where the mechanical mode interacts with the optical modes solely through dispersive couplings. This ideal framework allows us to clarify the key mechanism, i.e., the unconventional optical-mode hybridization, and to show how it can lead to an effective single-photon strong-coupling, sideband-resolved regime. In practice, however, microcavity optomechanical systems, such as the Fano-mirror setup considered here, can exhibit dissipative optomechanical couplings, as the mechanical displacement also modulates the optical decay rates, as briefly discussed in Sec.~\ref{SecModel}. This effect needs to be taken into account for realistic implementations. It can be incorporated by introducing position-dependent decay rates 
\begin{eqnarray}
\kappa_a(\hat{x}) &\approx& \kappa_{a,0} + \sqrt{2}g_{a,0}^{\kappa}\hat{x}, \nonumber\\ \kappa_d(\hat{x}) &\approx& \kappa_{d,0} + \sqrt{2}g_{d,0}^{\kappa}\hat{x}, 
\label{gadk}
\end{eqnarray}
where $g_{a,0}^{\kappa}$ and $g_{d,0}^{\kappa}$ denote the single-photon dissipative coupling strengths between the mechanical mode and the optical modes $a$ and $d$, respectively. In principle, one should also define $\gamma_{a}(\hat{x})$ in a similar manner for the right normal mirror. However, its effect can be absorbed into a slight renormalization of  $g_{a,0}^{\kappa}$.

\begin{figure}
\centering
\includegraphics[width = 0.55\linewidth]{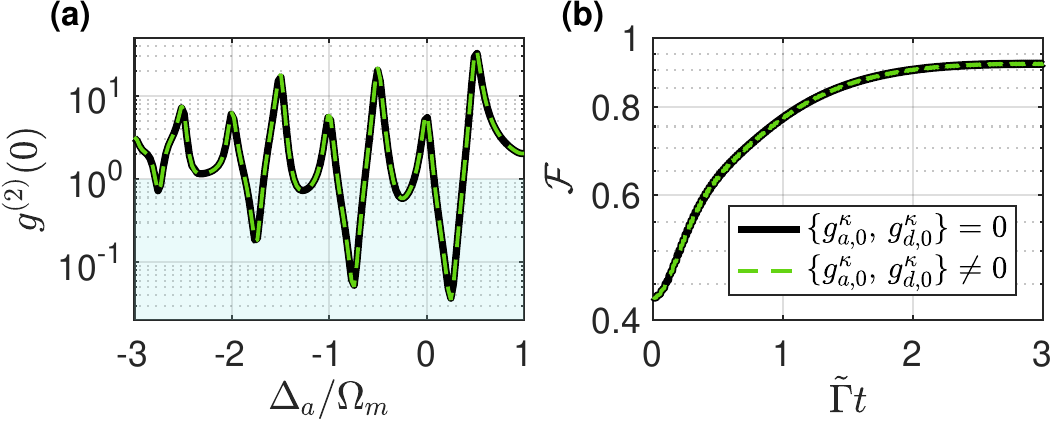}
\caption{Comparison of the results with and without the dissipative optomechanical couplings. We assume $g_{a,0}^{\kappa}/\Omega_{m}=0.25$ and $g_{d,0}^{\kappa}/\Omega_{m}=0.92$ for dissipative optomechanical couplings. In panel (a), we compare the second-order correlation function $g^{(2)}(0)$ of the $A_{-}$ mode, with all other parameters identical to those used for the purple dot-solid line in Fig.~\figpanel{Blockade}{a}. In panel (b), we compare the fidelity $\mathcal{F}$ of the prepared mechanical state with respect to an ideal even cat state, with all other parameters identical to those used for Fig.~\figpanel{CatS}{g}. The two panels share the same legend.}
\label{Dissipative}
\end{figure}

In this section, we investigate the impact of such dissipative optomechanical couplings on the results obtained above. This is achieved by replacing the constant optical decay rates in Eqs.~(\ref{La})--(\ref{Lad}) with the corresponding $x$-dependent counterparts in Eq.~(\ref{gadk}). We assume that the dissipative coupling strengths $g_{a,0}^{\kappa}$ and $g_{d,0}^{\kappa}$ are slightly larger in magnitude than the dispersive coupling strengths $g_{a,0}^{\omega}$ and $g_{d,0}^{\omega}$, yet remain of the same order of magnitude as in Ref.~\cite{WWoe2023}. In Fig.~\ref{Dissipative}, we compare the results obtained with and without including the dissipative optomechanical couplings. Specifically, in Figs.~\figpanel{Dissipative}{a} and \figpanel{Dissipative}{b}, the thick black solid lines reproduce the results in Figs.~\figpanel{Blockade}{a} (the purple dot-solid line) and \figpanel{CatS}{g}, respectively, while the thin red dashed lines represent the modified results for nonzero $g_{a,0}^{\kappa}$ and $g_{d,0}^{\kappa}$. In both panels, the excellent agreement between the two sets of results confirms that dissipative couplings of reasonable strengths have a negligible impact on the system’s behavior.

This agreement can be understood from two complementary perspectives. First, examining the $x$-dependent optical decay rates in Eq.~(\ref{gadk}), it is clear that the $x$-dependent terms are much smaller than the corresponding constant terms as long as $\{\kappa_{a,0},\kappa_{d,0}\}\gg\{g_{a,0}^{\kappa},g_{d,0}^{\kappa}\}$. Under this condition, the modulation of the decay rates due to the mechanical displacement is negligible, and thus has little influence on the dissipation processes. Second, from a more quantitative viewpoint, the \emph{effective} dissipative optomechanical coupling strengths for the two optical normal modes are given by $G_{\pm}^{\kappa}=\left[\partial\kappa_{\pm}(x)/\partial x\right]\vert_{x=0}/\sqrt{2}$~\cite{Juliette2023pra}, where $\kappa_{\pm}(x)$ are defined by replacing the constant decay rates in Eqs.~(\ref{keff}) and (\ref{keff2}) with their $x$-dependent counterparts. In the parameter regime we considered, where $\kappa_{a,0}\gg\kappa_{d,0}$ and the mixing angle $\theta$ is sufficiently small, one finds $G_{-}^{\kappa}\approx g_{d,0}^{\kappa}\left[1-\frac{1}{2}\sin{(2\theta)}\sqrt{\kappa_{a,0}/\kappa_{d,0}}\right]$, which is much smaller in magnitude than the dispersive counterpart $G_{-}\approx g_{d,0}^{\omega}$ provided that $|g_{d,0}^{\omega}|$ is comparable to $|g_{d,0}^{\kappa}|$. This again confirms that the effective dispersive coupling between the $A_{-}$ mode and the mechanical motion dominates the dynamics of the system. 

\subsection{Standard and dressed-state master equations}\label{SecDSME}

In the present work, we adopt the standard local master-equation description, in which the photonic reservoir induces dissipation only in the optical  degrees of freedom and the phononic reservoir induces dissipation only in the mechanical degrees of freedom. This description is appropriate for the parameter regimes considered in this paper (as will be verified) and offers a straightforward way to gain insight into how the hybridization of the optical modes facilitates the emergence of the above quantum phenomena. 

It has been shown that in the single-photon ultrastrong-coupling regime, a more accurate description is provided by the so-called ``dressed-state master equation''~\cite{HuDSME}, which formulates dissipation in the eigenbasis of the interacting optomechanical Hamiltonian. The dressed-state master equation introduces photon-number–dependent correction terms that generate additional mechanical damping and cavity dephasing. As the optomechanical coupling strength increases, these corrections cause the predictions of the dressed-state master equation to deviate increasingly from those of the standard description. 

Nevertheless, the correction terms scale with the mechanical damping rate $\gamma_{m}$. Therefore, when the mechanical damping is sufficiently weak (e.g., $\gamma_{m}\ll\kappa_{-}$ in our work), the difference between the two descriptions becomes negligible. In Appendix~\ref{AppD}, we verify this statement with a direct numerical comparison between the results obtained from the standard master equation and the dressed-state master equation. 

\subsection{Experimental platforms}

The presented Fano-mirror optomechanical setup could be realized with a suspended PhC mirror \cite{FanGuidedResonanceAnalysis2002} on top of a distributed Bragg reflector (DBR) mirror forming an optomechanical microcavity; see, e.g., recent works~\cite{WWoe2023,EnzianOE2023,MitraOE2024}. To obtain parameters similar to those assumed in Fig.~\ref{Blockade}, we set $\kappa_{a}=2\pi\times 1\,$THz, which (when choosing the same ratios $\omega_a/\kappa_a$ and $\omega_d/\kappa_a$ as in Fig.~\ref{Blockade}) results in $\omega_a$ and $\omega_d$ close to $2\pi \times 200$\,THz, i.e., corresponding to optical wavelengths in the telecom regime around 1500\,nm. The mechanical resonator will then have a frequency of $\Omega_{m}=2\pi \times 2\,$MHz and quality factor $Q_m=\Omega_{m}/\gamma_m=2.5\times10^7$, which is achievable with, for example, defect modes localized in membrane phononic crystals~\cite{TsaturyanNNano2017,CiersAPL2025}. To obtain $g_{a,0}^{\omega}/\Omega_{m}=0.25$, one could realize a microcavity with $L_\text{cav}\leq 1\,\mu$m~\cite{WWoe2023} and a mechanical defect mode with a mass $m_{\text{eff}}=1\,$ng~\cite{SaarinenOptica2023,EnzianOE2023}, such that $g_{a,0}^{\omega}=x_\text{zpf}\omega_a/L_\text{cav}\geq 0.2 \Omega_m$ with $x_\text{zpf}=\sqrt{\hbar/2m_{\text{eff}}\Omega_m}$. The suspended PhC provides the required localized optical mode $d$ and, thus, the left mirror. The linewidth of this PhC, i.e., the left mirror, would be $\kappa_d=2\pi \times 1.6\,$GHz, which translates into a quality factor of $10^5$ or a reflectivity of about 0.9999. PhC quality factors larger than $10^5$ can be achieved~\cite{YuriPRL2018HighQ,HuangNC2023HighQ,ZhouNL2025HighQ}, and recently a PhC reflectivity of $>0.9998$ has been demonstrated~\cite{ZhouLPR2023HighR}, close to the required value. The loss rate through the right mirror, $\gamma_a=2\pi\times 0.1\,$GHz, translates to a mirror reflectivity of larger than 0.99999, which is also experimentally achievable~\cite{RempeOL1992}. The optomechanical coupling between the PhC mode and the mechanical mode, $g_{d,0}^{\omega}$, as well as the coupling $\Lambda$ between the optical modes $a$ and $d$ would require careful analysis and engineering, which could be based on Ref.~\cite{WWoe2023}. Hence, the setup of a suspended PhC mirror forming a microcavity with a DBR constitutes a promising route to observe the analyzed nonclassical optomechanical states.

Alternative optomechanical platforms to realize similar dynamics could be based on coupled-mode cavity optomechanical~\cite{FangNP2017,BurgwalNC2023} or microwave electromechanical systems~\cite{YoussefiNature2022}. Together with the concept of reservoir engineering~\cite{MetelmannPRX2015}, these platforms could allow engineering of proper couplings between the optical and mechanical modes as well as of their couplings to the environment.

In realistic experiments, preparing a MHz-frequency mechanical oscillator in the quantum regime typically requires additional techniques, such as sideband cooling~\cite{Nunnenkamp2012Cooling,TransducerCooling2022} or measurement-based quantum control~\cite{MBF5,MHzCoolingKippenberg}. In particular, sideband cooling has also been extended to the single-photon strong-coupling regime~\cite{Nunnenkamp2012Cooling}. Moreover, it has been demonstrated that applying a cat-state preparation protocol to an initial thermal state can still produce a nonclassical ``hot Schr\"{o}dinger cat state''~\cite{HotCatState2025}, i.e., a coherent superposition of displaced thermal states. Remarkably, such hot cat states can retain clear quantum signatures (e.g., interference fringes and even Wigner-function negativity) despite having very low initial purity.

\section{Conclusions}

In summary, we have investigated a Fano-mirror microcavity optomechanical system in the quantum nonlinear regime, where the single-photon optomechanical coupling strengths are comparable to the mechanical frequency. The localized optical mode supported by the PhC membrane (i.e., the Fano mirror) and the Fabry-P\'{e}rot cavity mode exhibit an unconventional hybridization via both coherent and dissipative couplings, such that the optical subsystem can be well captured by a pair of bright and dark normal modes. Importantly, the dark mode can simultaneously exhibit strongly suppressed effective optical loss and sizable dispersive optomechanical coupling, which enables an effective sideband-resolved, single-photon (ultra)strong-coupling regime even when the bare optical linewidths are far (more than five orders of magnitude) larger than the mechanical frequency. 

Building on an effective master-equation description, we predicted clear signatures of quantum nonlinearity accessible in experimentally realistic parameter ranges. These include photon blockade of the dark normal mode, with antibunching resonances arising from the optomechanically induced Kerr-type anharmonicity, and deterministic generation of nonclassical mechanical states, such as mechanical cat states under a bichromatic driving protocol. We further showed that the key signatures are robust against dissipative optomechanical couplings of realistic strengths and, for the parameter regime considered in this work, are also consistent with a more refined dressed-state master-equation description. Overall, our results establish the Fano-mirror architecture as a promising route to accessing single-photon optomechanical nonlinearities and efficient mechanical state engineering in microcavity geometries where optical loss would otherwise preclude such physics. This opens up a new platform for nonclassical state generation and quantum technologies with continuous variables.

\section*{Data availability}
Data and computation code are available from the authors upon reasonable request.

\section*{Acknowledgments}
We thank B.~Hauer, H.~Pfeifer, Y.-T.~Chen, A.~Ciers, A.~Jung, G.~Ferrini, Y.~Li, T.~Huang, and F.~Zou for helpful discussions. We acknowledge financial support by the Knut och Alice Wallenberg stiftelse through project Grant No. 2022.0090, as well as through individual Wallenberg Academy fellowship (J.S.) and Scholar grants (W.W.).


\section*{Competing interests}
The authors declare no competing interests.

\appendix

\begin{widetext}
\section{Lindblad dissipators}

In this Appendix, we present different forms of the master equation~\eqref{MEexact}, first putting it in the canonical Lindblad form in Appendix~\ref{AppA:Lindblad} and, second, rewriting it in the normal mode basis in Appendix~\ref{AppA}.

\subsection{Canonical Lindblad form}\label{AppA:Lindblad}

As written, the dissipative part of the master equation in Eq.~\eqref{MEexact} does not make it immediately clear whether it represents a physically well-defined dynamics, since it is not expressed in canonical Lindblad form~\cite{Breuer2007Jan}. By contrast, $\mathcal{L}_{b}[\hat{\rho}]$, given in Eq.~\eqref{Lb}, is already in canonical form. We therefore focus exclusively on the optical dissipation, 
\begin{align}
    \mathcal{L}_\text{opt}[\hat{\rho}] &= \mathcal{L}_{a}[\hat{\rho}]+\mathcal{L}_{d}[\hat{\rho}]+\mathcal{L}_{ad}[\hat{\rho}] = \sum_{i, j} \Gamma_{ij} \left(\hat{c}_i\hat{\rho}\hat{c}_j^\dagger  - \frac{1}{2}\{\hat{c}_i^\dagger\hat{c}_j,\hat{\rho}\}\right),
\end{align}
where we have defined $\hat{c}_1 = \hat{a}$, $\hat{c}_2 = \hat{d}$ and the matrix
\begin{equation}
    \Gamma = \begin{pmatrix}
        \Gamma_a & \sqrt{\kappa_a\kappa_d}\\
        \sqrt{\kappa_a\kappa_d} & \kappa_d\\
    \end{pmatrix}.
\end{equation}
We now rewrite $\mathcal{L}_\text{opt}[\hat{\rho}]$ in canonical form by diagonalizing $\Gamma$, finding the eigenvalues $\Gamma_\pm$ associated with the eigenvectors $(x_1^\pm, x_2^\pm)^\text{T}$, with
\begin{gather}
    \Gamma_\pm = \frac{\Gamma_a + \kappa_d \pm \sqrt{(\Gamma_a + \kappa_d)^2 - 4\gamma_{a}\kappa_d}}{2},\label{Gam_pm} \\
    x_1^\pm = \frac{1}{\sqrt{N_\pm}}, \quad x_2^\pm = \frac{1}{2\sqrt{\kappa_a\kappa_d}\sqrt{N_\pm}}\left[(\kappa_d - \Gamma_{a}) \pm \sqrt{(\Gamma_a + \kappa_d)^2 - 4\gamma_{a}\kappa_d}\right],\\
    N_\pm = 1 + \frac{1}{4\kappa_a\kappa_d}\left[(\kappa_d - \Gamma_{a}) \pm \sqrt{(\Gamma_a + \kappa_d)^2 - 4\gamma_{a}\kappa_d}\right]^2.
\end{gather}
We thus obtain
\begin{equation}
     \mathcal{L}_\text{opt}[\hat{\rho}] = \sum_{\sigma=\pm} \Gamma_{\sigma} \left(\hat{c}_\sigma\hat{\rho}\hat{c}_\sigma^\dagger  - \frac{1}{2}\{\hat{c}_\sigma^\dagger\hat{c}_\sigma,\hat{\rho}\}\right),
\end{equation}
where $\hat{c}_\sigma = \sum_{j=1}^2 x_j^\sigma \hat{c}_j$ and
note that $\hat{c}_\pm$ obey the canonical bosonic commutation relations, namely $[\hat{c}_\sigma, \hat{c}_{\sigma'}^\dagger] = \delta_{\sigma\sigma'}$.
Finally, we can see from Eq.~\eqref{Gam_pm} that $\Gamma_\pm > 0$, and therefore, the master equation \eqref{MEexact} is indeed a Markovian Lindblad equation which preserves the Hermiticity and positivity of the system's density matrix.

\subsection{Lindblad dissipators in the optical normal-mode basis}\label{AppA}

In this Appendix, we explicitly derive the Lindblad dissipators in the optical normal-mode basis. After moving to this basis, the Lindblad dissipators in Eqs.~(\ref{La}) and (\ref{Ld}) become
\begin{eqnarray}
\tilde{\mathcal{L}}_a[\hat{\rho}] &=& \Gamma_a \left( \cos\theta \, \hat{A}_+ - \sin\theta \, \hat{A}_- \right) \hat{\rho} \left( \cos\theta \, \hat{A}_+^\dagger - \sin\theta \, \hat{A}_-^\dagger \right) \nonumber\\
&& - \frac{\Gamma_a}{2} \left\{ \left(\cos\theta \, \hat{A}_+^\dagger - \sin\theta \, \hat{A}_-^\dagger \right) \left(\cos\theta \, \hat{A}_+ - \sin\theta \hat{A}_- \right), \hat{\rho} \right\}, \label{Latilde}\\
\tilde{\mathcal{L}}_d[\hat{\rho}] &=& \kappa_d \left( \sin\theta \, \hat{A}_+ + \cos\theta \, \hat{A}_- \right) \hat{\rho} \left( \sin\theta \, \hat{A}_+^\dagger + \cos\theta \, \hat{A}_-^\dagger \right) \nonumber\\
&& - \frac{\kappa_d}{2} \left\{ \left(\sin\theta \, \hat{A}_+^\dagger + \cos\theta \, \hat{A}_-^\dagger \right) \left(\sin\theta \, \hat{A}_+ + \cos\theta \hat{A}_- \right), \hat{\rho} \right\}, \label{Ldtilde}
\end{eqnarray}
and the dissipator in Eq.~(\ref{Lad}) becomes
\begin{align}
\tilde{\mathcal{L}}_{ad}[\hat{\rho}] &= \sqrt{\kappa_a\kappa_d} \left( \cos\theta \, \hat{A}_+ - \sin\theta \, \hat{A}_- \right) \hat{\rho} \left( \sin\theta \, \hat{A}_+^\dagger + \cos\theta \, \hat{A}_-^\dagger \right) \nonumber\\
&\quad - \frac{\sqrt{\kappa_a\kappa_d}}{2} \left\{ \left(\cos\theta \, \hat{A}_+^\dagger - \sin\theta \, \hat{A}_-^\dagger \right)\left(\sin\theta \, \hat{A}_+ + \cos\theta \hat{A}_- \right), \hat{\rho} \right\} + \mathrm{H.c.}.
\label{Ladtilde}
\end{align}
Therefore, each normal mode $\hat{A}_\pm$ inherits dissipative contributions from both modes $a$ and $d$. The cross terms, such as $\hat{A}_+^\dagger \hat{A}_-$, give rise to cooperative dynamics between these two normal modes. 

Then, the master equation~\eqref{MEexact} can be rewritten as
\begin{equation}
    \dot{\hat{\rho}}=-i\left[\hat{\tilde{H}}_{\mathrm{tot}},\hat{\rho}\right]+\sum_{j=\{A_+,A_-,b,A_+A_-\}}\mathcal{L}_{j}[\hat{\rho}],
\end{equation}
with
\begin{subequations} \label{Lterms +-}
    \begin{align}
        \hat{\tilde{H}}_{\mathrm{tot}} &= \hat{\tilde{H}}_{\mathrm{opt}} + \hat{H}_{\mathrm{mec}} + \hat{\tilde{H}}_{\mathrm{om}}, \\
        \mathcal{L}_{A_\pm}[\hat{\rho}] &= \kappa_{\pm} \left( \hat{A}_\pm\hat{\rho} \hat{A}_\pm^\dagger - \frac{1}{2}\left\{ \hat{A}_\pm^\dagger \hat{A}_\pm, \hat{\rho} \right\} \right),\\
        \mathcal{L}_{A_+A_-}[\hat{\rho}] &=\left[\cos(2\theta) \sqrt{\kappa_a\kappa_d}  + \sin(2\theta)\frac{\kappa_d - \kappa_a}{2}\right] \left[ \hat{A}_+\hat{\rho}\hat{A}_-^\dagger - \frac{1}{2}\left\{ \hat{A}_+^\dagger\hat{A}_-, \hat{\rho} \right\} + \mathrm{H.c.} \right],
    \end{align}
\end{subequations}
where $\kappa_{\pm}$ are given by Eqs.~\eqref{keff} and \eqref{keff2}, and the components of $\hat{\tilde{H}}_{\mathrm{tot}}$ are given in Eqs.~\eqref{baremec}, \eqref{Hopttilde}, and \eqref{Homtilde}. When neglecting the contributions of mode $A_+$, the only relevant optical dissipator is $\mathcal{L}_{A_-}[\hat{\rho}]$, as stated around Eq.~\eqref{L-} in the main text.

\end{widetext}

\section{Effective parameters obtained from two different methods}\label{AppB}

\begin{figure}
\centering
\includegraphics[width = 0.55\linewidth]{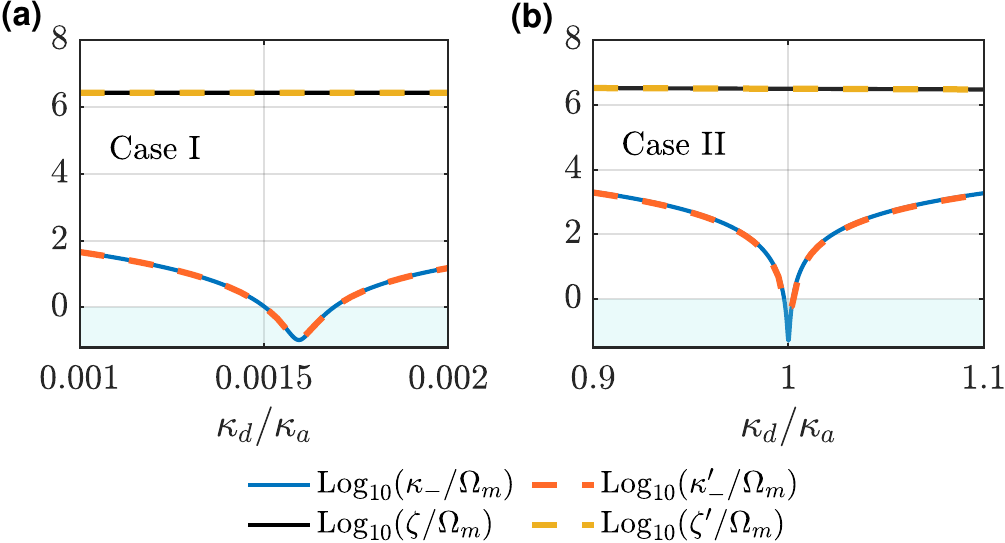}
\caption{Effective parameters obtained from both the effective master equation method (solid lines) and the effective Langevin equation method (dashed lines). In panel (a) we use identical parameters as those in Fig.~\figpanel{Eliminate}{a}, while in panel (b), we consider a different set of parameters (also considered in Ref.~\cite{Juliette2023pra}): $\gamma_a/\kappa_a = 6.4\times10^{-8}$, $\omega_a/\kappa_a = \omega_d/\kappa_a = 200$, $\Lambda/\kappa_a = 2$, and $\Omega_{m}/\kappa_a = 6.38\times10^{-7}$.}
\label{EffectP}
\end{figure}

As noted in the main text, the effective parameters of the two optical normal modes can also be determined within a Langevin-equation framework. By tracing out the environmental degrees of freedom and neglecting the weak mechanical contributions, one obtains the following Langevin equations for the optical modes: 
\begin{equation}
\begin{bmatrix}
\dot{\hat{a}} \\
\dot{\hat{d}}
\end{bmatrix}
= -i
\begin{bmatrix}
\omega_a - i\frac{\Gamma_a}{2} & \mathcal{G} \\
\mathcal{G} & \omega_d - i\frac{\kappa_d}{2}
\end{bmatrix}
\begin{bmatrix}
\hat{a} \\
\hat{d}
\end{bmatrix}
+ \vec{f}_{\mathrm{in}},
\label{Langevin}
\end{equation}
where $\vec{f}_{\mathrm{in}}$ is the vector of input fields. The non-Hermitian coefficient matrix in Eq.~(\ref{Langevin}) has two complex eigenvalues
\begin{equation}
\Omega_{\pm}' = \frac{\omega_a + \omega_d}{2} -i\frac{\Gamma_a + \kappa_d}{4} \pm  \sqrt{\left(\frac{\omega_{a}-\omega_{d}}{2} -i\frac{\Gamma_a - \kappa_d}{4} \right)^2 + \mathcal{G}^2},
\label{LEeigenvalues}
\end{equation}
whose real parts, $\omega_{\pm}'=\mathrm{Re}(\Omega_{\pm}')$, and imaginary parts, $\kappa_{\pm}'=-2\mathrm{Im}(\Omega_{\pm}')$, give the effective resonance frequencies and decay rates of the optical normal modes, respectively. In this framework, the frequency difference between the normal modes is given by $\Delta_{\mathrm{normal}}'=|\omega_{+}'-\omega_{-}'|$.

In the main text, we obtain the effective parameters from the effective master equation Eq.~(\ref{optME}). Here we validate this description by comparing them with those extracted from the Langevin-equation framework. 

We first recall the parameter set used in Fig.~\ref{Eliminate} and display the corresponding results in Fig.~\figpanel{EffectP}{a}. For visualization, we introduce $\zeta=\Delta_{\mathrm{normal}}-\kappa_{+}$ and $\zeta'=\Delta_{\mathrm{normal}}'-\kappa_{+}'$ as indicators of whether the $A_{+}$ mode can be safely eliminated. Positive values of $\zeta$ and $\zeta'$ ensure that a driving field resonant with the $A_-$ mode lies outside the spectral window of the $A_+$ mode. Larger positive values indicate a smaller dynamical contribution from $A_+$. The excellent consistency between the two sets of effective parameters validates the accuracy of the effective master equation approach adopted in the main text.

In fact, the parameter set used in Fig.~\figpanel{EffectP}{a} is not the only one that allows for safe elimination of the $A_+$ mode. In Fig.~\figpanel{EffectP}{b}, we examine an alternative set of parameters (also considered in Ref.~\cite{Juliette2023pra}), with which the two optical normal modes are well separated and the effective linewidth of the $A_-$ mode is much smaller than the mechanical frequency. In this regime, we again observe sufficiently good agreement between the results obtained from the two methods.

\section{Catability of the prepared cat states}\label{AppC}

To further characterize the generated mechanical cat states, we additionally evaluate them using the recently proposed notion of ``catability''~\cite{Catability2026}, which provides a complementary indicator to the fidelity and the Wigner logarithmic negativity used in the main text. While the fidelity quantifies the overlap with a chosen target cat state and the Wigner logarithmic negativity captures the nonclassicality of the prepared state without reference to a target state, catability is specifically designed to test whether a given state can be certified as an approximate cat state. For a given target cat-state size $\chi$ (determined by $\varepsilon_{p,1}$ in our scheme), catability can, in principle, be inferred from only three sets of displaced number-distribution measurements, rather than from a full reconstruction of the density matrix. This reduced measurement overhead makes catability an interesting tool for future experimental characterization of cat states.

For the even and odd cat states defined in Eq.~(\ref{catstates}),
the catability construction is based on the positive-semidefinite operator 
\begin{equation}
\hat O_{\pm}(\chi,\mu)
=
\left(\hat b^{\dagger 2}-\chi^{*2}\right)\left(\hat b^2-\chi^2\right)
+\mu\left(1\mp \hat \Pi\right),
\label{SPO}
\end{equation}
where $\hat \Pi$ is the parity operator, and $\mu>0$ is an auxiliary optimization parameter. The first term probes the separation of the two coherent peaks, while the parity term probes the quantum coherence between them (i.e., the fringes between the two peaks). For the even cat states considered in Fig.~\ref{CatS}, the relevant operator is $\hat{O}_{+}(\chi,\mu)$.

For a mechanical state $\hat{\rho}_b$, the corresponding even-cat catability is then defined as
\begin{equation}
\xi_{+}(\chi)
=
\min_{\mu>0}
\frac{
\mathrm{Tr}\!\left[\hat O_{+}(\chi,\mu)\hat\rho_b\right]
}{
\displaystyle \min_{\hat\rho_G}
\mathrm{Tr}\!\left[\hat O_{+}(\chi,\mu)\hat\rho_G\right]
},
\end{equation}
where the minimization in the denominator is taken over the set of all single-mode Gaussian states $\hat{\rho}_{G}$. According to Ref.~\cite{Catability2026}, the interpretation of this quantity is as follows: $\xi_{+}(\chi)=0$ corresponds to the ideal target cat state, $0<\xi_{+}(\chi)<1$ indicates that the tested state is a non-Gaussian approximate even cat state of size $\chi$, while $\xi_{+}(\chi)\ge 1$ makes the criterion inconclusive.

\begin{figure}
\centering
\includegraphics[width = 0.55\linewidth]{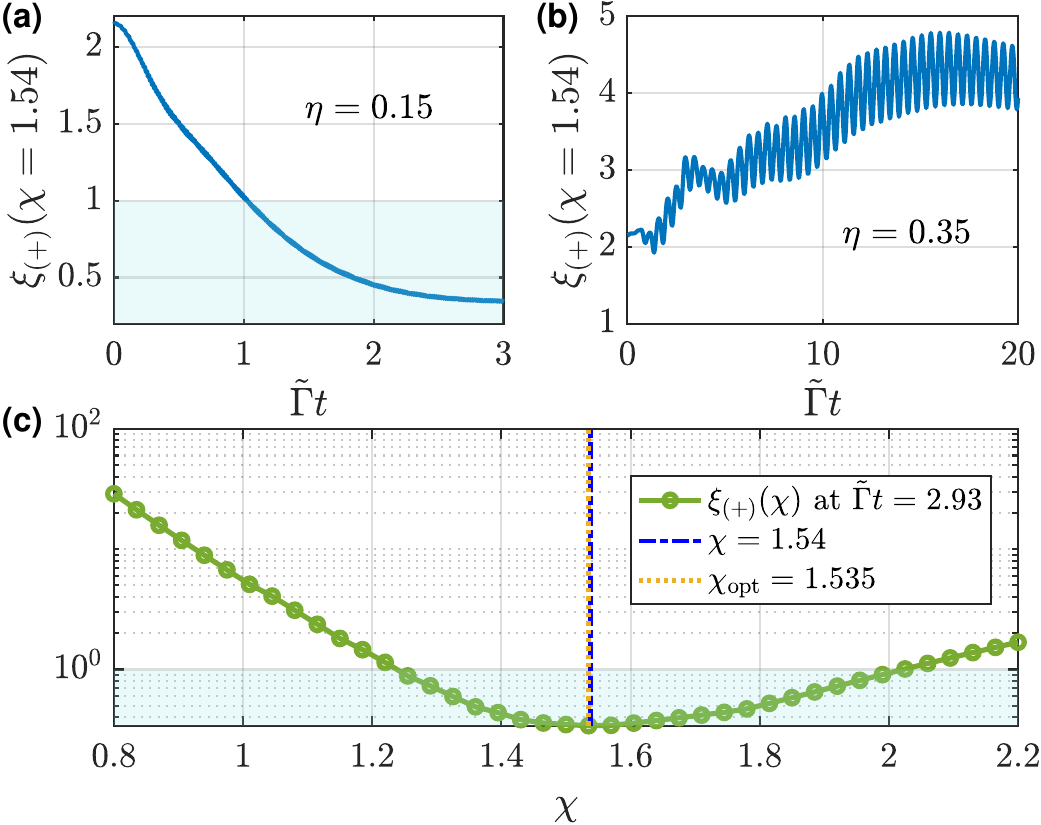}
\caption{Additional characterization of the mechanically prepared states via the catability criterion. (a) Time evolution of the even-cat catability $\xi_{(+)}(\chi)$ for the generated mechanical state at $\eta=0.15$, using the same cat size $\chi=1.54$ as in Fig.~\ref{CatS}. (b) Same as in (a), but for $\eta=0.35$. (c) Catability $\xi_{(+)}(\chi)$ at $\tilde{\Gamma}t=2.93$ for the $\eta=0.15$ case, plotted as a function of the cat size $\chi$. The minimum occurs at $\chi\simeq 1.535$, which is essentially identical to the target value $\chi=1.54$. The light blue areas in (a) and (c) represent the region of $\xi_{(+)}(\chi)<1$, where the state is demonstrated to be a non-Gaussian approximate even cat state. For the catability evaluation, we assume $\mu\in[0.05,5]$ with $15$ sampling points and $\chi\in[0.8,2.2]$ with $41$ sampling points. The minimization over Gaussian states is performed using $8$ multistart initial guesses. All other parameters are the same as in Fig.~\ref{CatS}.}
\label{CatabilityFig}
\end{figure}

In Figs.~\figpanel{CatabilityFig}{a} and \figpanel{CatabilityFig}{b}, we plot the time evolution of $\xi_{(+)}(\chi)$ for the generated mechanical state (using the same protocol as in Fig.~\ref{CatS}), fixing the cat size to the target value $\chi=1.54$ prescribed by our driving protocol (see also Fig.~\ref{CatS}). For the weaker-coupling case $\eta=0.15$ [panel (a)], the catability decreases monotonically from values above unity and eventually reaches values well below unity. This confirms that the prepared mechanical state becomes a non-Gaussian approximate even cat state. By contrast, for the stronger-coupling case $\eta=0.35$ [panel (b)], the catability remains above unity throughout the shown time interval. In this case, the catability does not identify the state as an approximate even cat of the chosen size, although it does not imply that the state is not catlike. 

In Fig.~\figpanel{CatabilityFig}{c}, we further evaluate $\xi_{(+)}(\chi)$ at the time $\tilde{\Gamma}t=2.93$ (corresponding to the maximum fidelity for the $\eta=0.15$ case; see Fig.~\ref{CatS}) as a function of the cat size $\chi$. The minimum is found at $\chi\simeq 1.535$, which is essentially identical to the target value $\chi=1.54$. This implies that the target cat state employed for the fidelity is already very close to the ideal even cat state that best matches the prepared mechanical state. Taken together, these results provide an additional consistency check for our mechanical state-preparation protocol.

\section{Comparison with the dressed-state master equation method}\label{AppD}

\begin{figure}
\centering
\includegraphics[width = 0.55\linewidth]{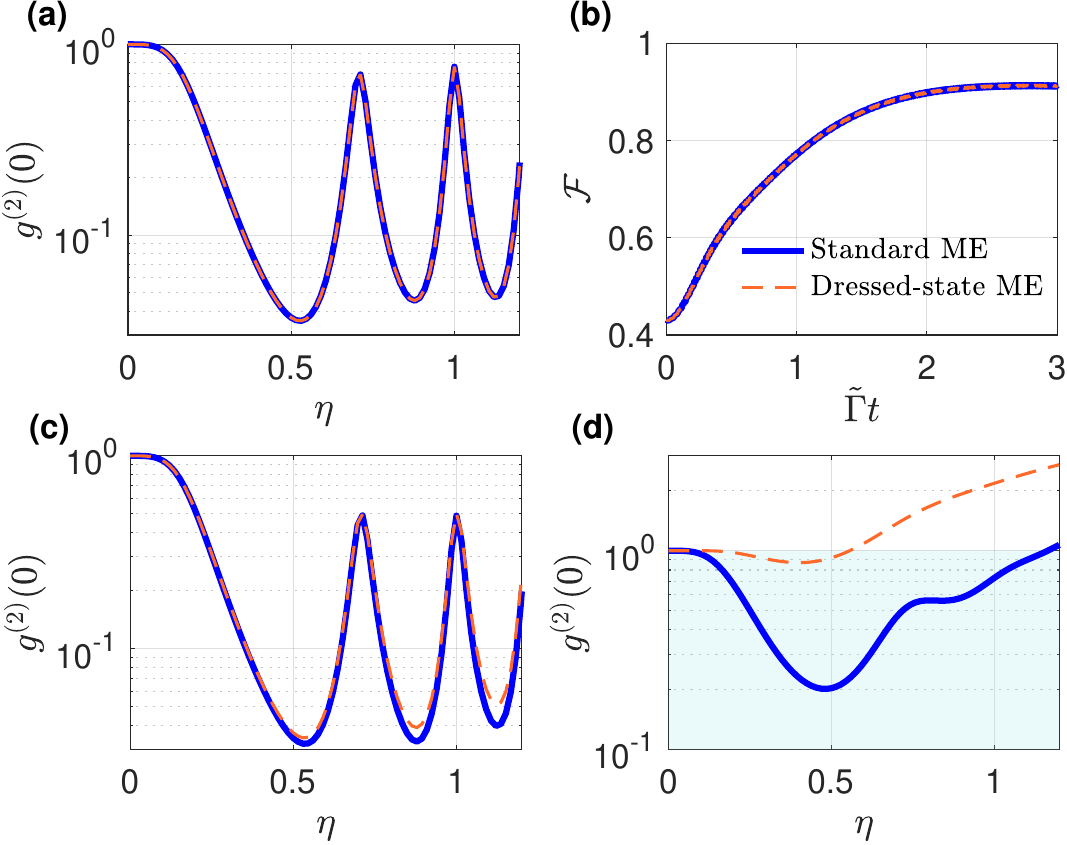}
\caption{Comparison of the results obtained from a standard master equation [cf. Eqs.~(\ref{MEeff}) and (\ref{MEcat})] and a dressed-state master equation [cf. Eq.~(\ref{eqDSME})]. In panel (a), we compare the second-order correlation function $g^{(2)}(0)$ of the $A_{-}$ mode, with all other parameters identical to those used for the purple dot-solid line in Fig.~\figpanel{Blockade}{a}. In panel (b), we compare the fidelity $\mathcal{F}$ of the prepared mechanical state with respect to an ideal even cat state, with all other parameters identical to those used for Fig.~\figpanel{CatS}{g}. Panels (c) and (d) examine the same quantities as in panel (a), but with larger mechanical damping rates: $\gamma_{m}/\kappa_a = 8 \times 10^{-9}$ (i.e., $\gamma_{m}/\Omega_{m}=4\times10^{-3}$) in (c) and $\gamma_{m}/\kappa_a = 8 \times 10^{-7}$ (i.e., $\gamma_{m}/\Omega_{m}=0.4$) in (d). All panels share the same legend as shown in (b).}
\label{DSME}
\end{figure}

Strictly speaking, the master equation of an optomechanical system should be appropriately modified when it is operated in the single-photon ultrastrong-coupling regime~\cite{HuDSME}. In this regime, the mechanical component of the optomechanical dressed states undergoes a significant \emph{displacement}, which is proportional to the photon number, due to the strong optomechanical interaction. Moreover, the system experiences an \emph{effective pure dephasing} of the optical mode, which arises from the mechanical noise mediated by the optomechanical coupling. Under the Born-Markov and the rotating-wave approximations, the modified master equation (referred to as the ``dressed-state master equation'') can be derived as~\cite{HuDSME}
\begin{eqnarray}
\dot{\hat{\hat{\rho}}} &=& -i[\hat{H},\hat{\hat{\rho}}]+\gamma_{m}(n_{\mathrm{th},b}+1)\mathcal{D}[\hat{b}-\eta\hat{n}_{-}]\hat{\hat{\rho}} + \gamma_{m}n_{\mathrm{th},b}\mathcal{D}[\hat{b}^{\dag}-\eta\hat{n}_{-}]\hat{\hat{\rho}} \nonumber\\
&&+\kappa_{-}\mathcal{D}[\hat{A}_{-}]\hat{\hat{\rho}}+4\gamma_{m}\left(k_{B}T/\Omega_{m}\right)\eta^{2}\mathcal{D}[\hat{n}_{-}]\hat{\hat{\rho}},
\label{eqDSME}
\end{eqnarray}
where $\hat{\hat{\rho}}$ and $\hat{H}$ denote the density matrix and Hamiltonian of interest, respectively, $k_{B}$ is the Boltzmann constant, and all other symbols are defined in the main text. Here the Lindblad superoperator is defined as $\mathcal{D}[\hat{o}]\hat{\hat{\rho}}=\hat{o}\hat{\hat{\rho}}\hat{o}^{\dag}-\frac{1}{2}\{\hat{o}^{\dag}\hat{o},\hat{\hat{\rho}}\}$, which, together with the corresponding decay rates, forms the superoperators $\mathcal{L}_{o}$ introduced in the main text. 
Clearly, Eq.~(\ref{eqDSME}) reduces to the standard form in the regime of $\eta=-G_{-}/\Omega_{m}\ll1$.

In Fig.~\ref{DSME}, we present the results obtained from the standard master equation [used in the main text; cf. Eqs.~(\ref{MEeff}) and (\ref{MEcat})] and the dressed-state master equation [cf. Eq.~(\ref{eqDSME})]. Specifically, Figs~\figpanel{DSME}{a} and \figpanel{DSME}{b} examine the same quantities as in Figs.~\figpanel{Dissipative}{a} and \figpanel{Dissipative}{b}, respectively. The excellent coincidence between the two approaches appears to contradict the findings in Ref.~\cite{HuDSME}, which predict a pronounced discrepancy in the ultra-strong coupling regime. The coincidence in Fig.~\ref{DSME} can be understood from the fact that we consider a sufficiently small mechanical damping rate $\gamma_{m}$ (over three orders of magnitude smaller than $\kappa_{-}$), which determines the magnitudes of the correction terms in Eq.~(\ref{eqDSME}). Such a small $\gamma_{m}$ is well within reach in current experiments~\cite{WWoe2023}. To verify this explanation, we show in Figs.~\figpanel{DSME}{c} and \figpanel{DSME}{d} the same comparison as in Fig.~\figpanel{DSME}{a}, but for two much larger mechanical damping rates [$\gamma_{m}/\Omega_{m}=4\times10^{-3}$ in Fig.~\figpanel{DSME}{c} and $\gamma_{m}/\Omega_{m}=0.4$ in Fig.~\figpanel{DSME}{d}]. Clear discrepancies emerge as $\eta$ increases, and they become progressively more pronounced for larger $\gamma_m$. This indicates that under strong mechanical damping, the dressed-state master equation typically predicts more classical behavior than the standard master equation.

\bibliography{OMRefs}

@book{Breuer2007Jan,
    title = {The {{Theory}} of {{Open Quantum Systems}}},
    author = {Breuer, Heinz-Peter and Petruccione, Francesco},
    year = 2007,
    month = jan,
    edition = {1},
    publisher = {Oxford University PressOxford},
    doi = {10.1093/acprof:oso/9780199213900.001.0001},
    urldate = {2026-01-16},
    isbn = {978-0-19-921390-0 978-0-19-170634-9},
    langid = {english}
}

@article{OM2014RMP,
  title = {Cavity optomechanics},
  author = {Aspelmeyer, Markus and Kippenberg, Tobias J. and Marquardt, Florian},
  journal = {Rev. Mod. Phys.},
  volume = {86},
  issue = {4},
  pages = {1391--1452},
  numpages = {62},
  year = {2014},
  month = {Dec},
  publisher = {American Physical Society},
  doi = {10.1103/RevModPhys.86.1391},
  url = {https://link.aps.org/doi/10.1103/RevModPhys.86.1391}
}

@article{OMmeasure1,
  title = {{Measuring nanomechanical motion with a microwave cavity interferometer}},
  author = {Regal, C. A. and Teufel, J. D. and Lehnert, K. W.},
  journal = {Nat. Phys.},
  volume = {4},
  issue = {7},
  pages = {555},
  year = {2008},
  month = {May},
  publisher = {Nature Publication Group},
  doi = {10.1038/nphys974},
  url = {https://www.nature.com/articles/nphys974#citeas}
}

@article{OMmeasure2,
  title = {{Resolved-sideband cooling and position measurement of a micromechanical oscillator close to the Heisenberg uncertainty limit}},
  author = {Schliesser, Albert and Arcizet, Olivier and Rivi{\`e}re, R{\'e}mi and Anetsberger, Georg and Kippenberg, Tobias J},
  journal = {Nat. Phys.},
  volume = {5},
  number = {7},
  pages = {509},
  year = {2009},
  publisher = {Nature Publishing Group},
  doi = {10.1038/nphys1304},
  url = {https://www.nature.com/articles/nphys1304}
}

@article{OMmeasure3,
  title={{A hybrid on-chip optomechanical transducer for ultrasensitive force measurements}},
  author={Gavartin, Emanuel and Verlot, Pierre and Kippenberg, Tobias J},
  journal={Nat. Nanotechnol.},
  volume={7},
  number={8},
  pages={509},
  year={2012},
  publisher={Nature Publishing Group},
  doi = {10.1038/nnano.2012.97},
  url = {https://www.nature.com/articles/nnano.2012.97}
}

@article{OMQT,
  title={Optomechanics for quantum technologies},
  author={Barzanjeh, Shabir and Xuereb, Andr{\'e} and Gr{\"o}blacher, Simon and Paternostro, Mauro and Regal, Cindy A and Weig, Eva M},
  journal={Nat. Phys.},
  volume={18},
  number={1},
  pages={15--24},
  year={2022},
  publisher={Nature Publishing Group},
  doi = {10.1038/s41567-021-01402-0},
  url = {https://www.nature.com/articles/s41567-021-01402-0}
}

@article{MBF5,
  title={Measurement-based quantum control of mechanical motion},
  author={Rossi, Massimiliano and Mason, David and Chen, Junxin and Tsaturyan, Yeghishe and Schliesser, Albert},
  journal={Nature},
  volume={563},
  number={7729},
  pages={53--58},
  year={2018},
  publisher={Nature Publishing Group},
  doi = {10.1038/s41586-018-0643-8},
  url = {https://www.nature.com/articles/s41586-018-0643-8}
}

@article{2cavity1,
  title = {Electromagnetically-induced-transparency-like ground-state cooling in a double-cavity optomechanical system},
  author = {Guo, Yujie and Li, Kai and Nie, Wenjie and Li, Yong},
  journal = {Phys. Rev. A},
  volume = {90},
  issue = {5},
  pages = {053841},
  numpages = {6},
  year = {2014},
  month = {Nov},
  publisher = {American Physical Society},
  doi = {10.1103/PhysRevA.90.053841},
  url = {https://link.aps.org/doi/10.1103/PhysRevA.90.053841}
}

@article{2cavity2,
author = {Yang, Jun-Ya and Wang, Dong-Yang and Bai, Cheng-Hua and Guan, Si-Yu and Gao, Xiao-Yuan and Zhu, Ai-Dong and Wang, Hong-Fu},
journal = {Opt. Express},
number = {16},
pages = {22855--22867},
publisher = {Optica Publishing Group},
title = {Ground-state cooling of mechanical oscillator via quadratic optomechanical coupling with two coupled optical cavities},
volume = {27},
month = {Aug},
year = {2019},
url = {https://opg.optica.org/oe/abstract.cfm?URI=oe-27-16-22855},
doi = {10.1364/OE.27.022855},
}

@article{2cavity3,
  title={Cavity-assisted coherent feedback cooling of a mechanical resonator to the ground-state in the unresolved sideband regime},
  author={Mansouri, Daryoosh and Rezaie, Behrooz and Ranjbar, Abolfazl and Daeichian, Abolghasem},
  journal={J. Phys. B},
  volume={55},
  number={16},
  pages={165501},
  year={2022},
  publisher={IOP Publishing},
  doi = {10.1088/1361-6455/ac7d27},
  url = {https://iopscience.iop.org/article/10.1088/1361-6455/ac7d27#bac7d27s6}
}

@article{FanoM,
  title = {Cavity Quantum Electrodynamics with Frequency-Dependent Reflectors},
  author = {{\v{C}}ernot{\'\i}k, Ond{\v{r}}ej and Dantan, Aur\'elien and Genes, Claudiu},
  journal = {Phys. Rev. Lett.},
  volume = {122},
  issue = {24},
  pages = {243601},
  numpages = {7},
  year = {2019},
  month = {Jun},
  publisher = {American Physical Society},
  doi = {10.1103/PhysRevLett.122.243601},
  url = {https://link.aps.org/doi/10.1103/PhysRevLett.122.243601}
}

@article{Juliette2021pra,
  title = {Optomechanical cooling with coherent and squeezed light: The thermodynamic cost of opening the heat valve},
  author = {Monsel, Juliette and Dashti, Nastaran and Manjeshwar, Sushanth Kini and Eriksson, Jakob and Ernbrink, Henric and Olsson, Ebba and Torneus, Emelie and Wieczorek, Witlef and Splettstoesser, Janine},
  journal = {Phys. Rev. A},
  volume = {103},
  issue = {6},
  pages = {063519},
  numpages = {25},
  year = {2021},
  month = {Jun},
  publisher = {American Physical Society},
  doi = {10.1103/PhysRevA.103.063519},
  url = {https://link.aps.org/doi/10.1103/PhysRevA.103.063519}
}

@article{WWoe2023,
author = {Manjeshwar, Sushanth Kini and Ciers, Anastasiia and Monsel, Juliette and Pfeifer, Hannes and Peralle, Cindy and Wang, Shu Min and Tassin, Philippe and Wieczorek, Witlef},
journal = {Opt. Express},
number = {19},
pages = {30212--30226},
publisher = {Optica Publishing Group},
title = {{Integrated microcavity optomechanics with a suspended photonic crystal mirror above a distributed Bragg reflector}},
volume = {31},
month = {Sep},
year = {2023},
url = {https://opg.optica.org/oe/abstract.cfm?URI=oe-31-19-30212},
doi = {10.1364/OE.496447},
}

@article{SVCR2,
  title = {Single-Photon Optomechanics},
  author = {Nunnenkamp, A. and B\o{}rkje, K. and Girvin, S. M.},
  journal = {Phys. Rev. Lett.},
  volume = {107},
  issue = {6},
  pages = {063602},
  numpages = {5},
  year = {2011},
  month = {Aug},
  publisher = {American Physical Society},
  doi = {10.1103/PhysRevLett.107.063602},
  url = {https://link.aps.org/doi/10.1103/PhysRevLett.107.063602}
}

@article{Juliette2023pra,
    title = {Dissipative and Dispersive Cavity Optomechanics with a Frequency-Dependent Mirror},
    author = {Monsel, Juliette and Ciers, Anastasiia and Manjeshwar, Sushanth Kini and Wieczorek, Witlef and Splettstoesser, Janine},
    year = {2024},
    month = apr,
    journal = {Phys. Rev. A},
    volume = {109},
    number = {4},
    pages = {043532},
    publisher = {{American Physical Society}},
    doi = {10.1103/PhysRevA.109.043532}
}

@article{Harris2008nature,
  title={Strong dispersive coupling of a high-finesse cavity to a micromechanical membrane},
  author={Thompson, JD and Zwickl, BM and Jayich, AM and Marquardt, Florian and Girvin, SM and Harris, JGE},
  journal={Nature},
  volume={452},
  number={7183},
  pages={72--75},
  year={2008},
  publisher={Nature Publishing Group},
  doi = {10.1038/nature06715},
  url = {https://www.nature.com/articles/nature06715}
}

@article{YCLiu2015pra,
  title = {Coupled cavities for motional ground-state cooling and strong optomechanical coupling},
  author = {Liu, Yong-Chun and Xiao, Yun-Feng and Luan, Xingsheng and Gong, Qihuang and Wong, Chee Wei},
  journal = {Phys. Rev. A},
  volume = {91},
  issue = {3},
  pages = {033818},
  numpages = {10},
  year = {2015},
  month = {Mar},
  publisher = {American Physical Society},
  doi = {10.1103/PhysRevA.91.033818},
  url = {https://link.aps.org/doi/10.1103/PhysRevA.91.033818}
}

@article{Hauer2023prl,
  title = {Nonlinear Sideband Cooling to a Cat State of Motion},
  author = {Hauer, B. D. and Combes, J. and Teufel, J. D.},
  journal = {Phys. Rev. Lett.},
  volume = {130},
  issue = {21},
  pages = {213604},
  numpages = {7},
  year = {2023},
  month = {May},
  publisher = {American Physical Society},
  doi = {10.1103/PhysRevLett.130.213604},
  url = {https://link.aps.org/doi/10.1103/PhysRevLett.130.213604}
}

@article{Wigner1932,
  title = {On the Quantum Correction For Thermodynamic Equilibrium},
  author = {Wigner, E.},
  journal = {Phys. Rev.},
  volume = {40},
  issue = {5},
  pages = {749--759},
  numpages = {0},
  year = {1932},
  month = {Jun},
  publisher = {American Physical Society},
  doi = {10.1103/PhysRev.40.749},
  url = {https://link.aps.org/doi/10.1103/PhysRev.40.749}
}

@article{DL2025Fano,
  title = {Coherent feedback control for cavity optomechanical systems with a frequency-dependent mirror},
  author = {Du, Lei and Monsel, Juliette and Wieczorek, Witlef and Splettstoesser, Janine},
  journal = {Phys. Rev. A},
  volume = {111},
  issue = {1},
  pages = {013506},
  numpages = {17},
  year = {2025},
  month = {Jan},
  publisher = {American Physical Society},
  doi = {10.1103/PhysRevA.111.013506},
  url = {https://link.aps.org/doi/10.1103/PhysRevA.111.013506}
}

@article{RablPB2011,
  title = {Photon Blockade Effect in Optomechanical Systems},
  author = {Rabl, P.},
  journal = {Phys. Rev. Lett.},
  volume = {107},
  issue = {6},
  pages = {063601},
  numpages = {5},
  year = {2011},
  month = {Aug},
  publisher = {American Physical Society},
  doi = {10.1103/PhysRevLett.107.063601},
  url = {https://link.aps.org/doi/10.1103/PhysRevLett.107.063601}
}

@article{Teufel2011Nature,
  title = {{Circuit cavity electromechanics in the strong-coupling regime}},
  volume = {471},
  ISSN = {1476-4687},
  url = {http://dx.doi.org/10.1038/nature09898},
  DOI = {10.1038/nature09898},
  number = {7337},
  journal = {Nature},
  publisher = {Springer Science and Business Media LLC},
  author = {Teufel,  J. D. and Li,  Dale and Allman,  M. S. and Cicak,  K. and Sirois,  A. J. and Whittaker,  J. D. and Simmonds,  R. W.},
  year = {2011},
  month = mar,
  pages = {204–208}
}

@article{Chan2011Nature,
  title = {{Laser cooling of a nanomechanical oscillator into its quantum ground state}},
  volume = {478},
  ISSN = {1476-4687},
  url = {http://dx.doi.org/10.1038/nature10461},
  DOI = {10.1038/nature10461},
  number = {7367},
  journal = {Nature},
  publisher = {Springer Science and Business Media LLC},
  author = {Chan,  Jasper and Alegre,  T. P. Mayer and Safavi-Naeini,  Amir H. and Hill,  Jeff T. and Krause,  Alex and Gr\"{o}blacher,  Simon and Aspelmeyer,  Markus and Painter,  Oskar},
  year = {2011},
  month = oct,
  pages = {89–92}
}

@article{Leijssen2017NC,
  title = {{Nonlinear cavity optomechanics with nanomechanical thermal fluctuations}},
  volume = {8},
  ISSN = {2041-1723},
  url = {http://dx.doi.org/10.1038/ncomms16024},
  DOI = {10.1038/ncomms16024},
  number = {1},
  journal = {Nat. Commun.},
  publisher = {Springer Science and Business Media LLC},
  author = {Leijssen,  Rick and La Gala,  Giada R. and Freisem,  Lars and Muhonen,  Juha T. and Verhagen,  Ewold},
  year = {2017},
  month = jul,
  pages = {16024}
}

@article{HuDSME,
  title = {Quantum coherence in ultrastrong optomechanics},
  author = {Hu, Dan and Huang, Shang-Yu and Liao, Jie-Qiao and Tian, Lin and Goan, Hsi-Sheng},
  journal = {Phys. Rev. A},
  volume = {91},
  issue = {1},
  pages = {013812},
  numpages = {8},
  year = {2015},
  month = {Jan},
  publisher = {American Physical Society},
  doi = {10.1103/PhysRevA.91.013812},
  url = {https://link.aps.org/doi/10.1103/PhysRevA.91.013812}
}

@article{LiaoUSC2020,
  title = {Generalized ultrastrong optomechanical-like coupling},
  author = {Liao, Jie-Qiao and Huang, Jin-Feng and Tian, Lin and Kuang, Le-Man and Sun, Chang-Pu},
  journal = {Phys. Rev. A},
  volume = {101},
  issue = {6},
  pages = {063802},
  numpages = {15},
  year = {2020},
  month = {Jun},
  publisher = {American Physical Society},
  doi = {10.1103/PhysRevA.101.063802},
  url = {https://link.aps.org/doi/10.1103/PhysRevA.101.063802}
}

@article{LudwigPRL2012,
  title = {{Enhanced Quantum Nonlinearities in a Two-Mode Optomechanical System}},
  author = {Ludwig, Max and Safavi-Naeini, Amir H. and Painter, Oskar and Marquardt, Florian},
  journal = {Phys. Rev. Lett.},
  volume = {109},
  issue = {6},
  pages = {063601},
  numpages = {5},
  year = {2012},
  month = {Aug},
  publisher = {American Physical Society},
  doi = {10.1103/PhysRevLett.109.063601},
  url = {https://link.aps.org/doi/10.1103/PhysRevLett.109.063601}
}

@article{NaseemGME,
  title = {Thermodynamic consistency of the optomechanical master equation},
  author = {Naseem, M. Tahir and Xuereb, Andr\'e and M\"ustecapl\ifmmode \imath \else \i \fi{}o\ifmmode \breve{g}\else \u{g}\fi{}lu, \"Ozg\"ur E.},
  journal = {Phys. Rev. A},
  volume = {98},
  issue = {5},
  pages = {052123},
  numpages = {9},
  year = {2018},
  month = {Nov},
  publisher = {American Physical Society},
  doi = {10.1103/PhysRevA.98.052123},
  url = {https://link.aps.org/doi/10.1103/PhysRevA.98.052123}
}

@article{OConnell2010,
  title = {Quantum ground state and single-phonon control of a mechanical resonator},
  volume = {464},
  ISSN = {1476-4687},
  url = {http://dx.doi.org/10.1038/nature08967},
  DOI = {10.1038/nature08967},
  number = {7289},
  journal = {Nature},
  publisher = {Springer Science and Business Media LLC},
  author = {O’Connell,  A. D. and Hofheinz,  M. and Ansmann,  M. and Bialczak,  Radoslaw C. and Lenander,  M. and Lucero,  Erik and Neeley,  M. and Sank,  D. and Wang,  H. and Weides,  M. and Wenner,  J. and Martinis,  John M. and Cleland,  A. N.},
  year = {2010},
  month = mar,
  pages = {697}
}

@article{Bose1997PRA,
  title = {Preparation of nonclassical states in cavities with a moving mirror},
  author = {Bose, S. and Jacobs, K. and Knight, P. L.},
  journal = {Phys. Rev. A},
  volume = {56},
  issue = {5},
  pages = {4175--4186},
  numpages = {0},
  year = {1997},
  month = {Nov},
  publisher = {American Physical Society},
  doi = {10.1103/PhysRevA.56.4175},
  url = {https://link.aps.org/doi/10.1103/PhysRevA.56.4175}
}

@article{Marshall2003PRL,
  title = {{Towards Quantum Superpositions of a Mirror}},
  author = {Marshall, William and Simon, Christoph and Penrose, Roger and Bouwmeester, Dik},
  journal = {Phys. Rev. Lett.},
  volume = {91},
  issue = {13},
  pages = {130401},
  numpages = {4},
  year = {2003},
  month = {Sep},
  publisher = {American Physical Society},
  doi = {10.1103/PhysRevLett.91.130401},
  url = {https://link.aps.org/doi/10.1103/PhysRevLett.91.130401}
}

@article{Garziano2015PRA,
  title = {Single-step arbitrary control of mechanical quantum states in ultrastrong optomechanics},
  author = {Garziano, L. and Stassi, R. and Macr\'{\i}, V. and Savasta, S. and Di Stefano, O.},
  journal = {Phys. Rev. A},
  volume = {91},
  issue = {2},
  pages = {023809},
  numpages = {7},
  year = {2015},
  month = {Feb},
  publisher = {American Physical Society},
  doi = {10.1103/PhysRevA.91.023809},
  url = {https://link.aps.org/doi/10.1103/PhysRevA.91.023809}
}

@article{Pinard1995PRA,
  title = {Quantum-nondemolition measurement of light by a piezoelectric crystal},
  author = {Pinard, M. and Fabre, C. and Heidmann, A.},
  journal = {Phys. Rev. A},
  volume = {51},
  issue = {3},
  pages = {2443},
  numpages = {0},
  year = {1995},
  month = {Mar},
  publisher = {American Physical Society},
  doi = {10.1103/PhysRevA.51.2443},
  url = {https://link.aps.org/doi/10.1103/PhysRevA.51.2443}
}

@article{Qvarfort2018NC,
  title = {Gravimetry through non-linear optomechanics},
  volume = {9},
  ISSN = {2041-1723},
  url = {http://dx.doi.org/10.1038/s41467-018-06037-z},
  DOI = {10.1038/s41467-018-06037-z},
  number = {1},
  journal = {Nat. Commun.},
  publisher = {Springer Science and Business Media LLC},
  author = {Qvarfort,  Sofia and Serafini,  Alessio and Barker,  P. F. and Bose,  Sougato},
  year = {2018},
  month = sep,
  pages = {3690}
}

@article{Groblacher2009Nature,
  title = {Observation of strong coupling between a micromechanical resonator and an optical cavity field},
  volume = {460},
  ISSN = {1476-4687},
  url = {http://dx.doi.org/10.1038/nature08171},
  DOI = {10.1038/nature08171},
  number = {7256},
  journal = {Nature},
  publisher = {Springer Science and Business Media LLC},
  author = {Gr\"{o}blacher,  Simon and Hammerer,  Klemens and Vanner,  Michael R. and Aspelmeyer,  Markus},
  year = {2009},
  month = aug,
  pages = {724}
}

@article{Clerk2011Kerr,
  title = {{Quantum-limited amplification with a nonlinear cavity detector}},
  author = {Laflamme, C. and Clerk, A. A.},
  journal = {Phys. Rev. A},
  volume = {83},
  issue = {3},
  pages = {033803},
  numpages = {13},
  year = {2011},
  month = {Mar},
  publisher = {American Physical Society},
  doi = {10.1103/PhysRevA.83.033803},
  url = {https://link.aps.org/doi/10.1103/PhysRevA.83.033803}
}

@article{Metelmann2023Kerr,
  title = {{Kerr Enhanced Backaction Cooling in Magnetomechanics}},
  author = {Zoepfl, D. and Juan, M. L. and Diaz-Naufal, N. and Schneider, C. M. F. and Deeg, L. F. and Sharafiev, A. and Metelmann, A. and Kirchmair, G.},
  journal = {Phys. Rev. Lett.},
  volume = {130},
  issue = {3},
  pages = {033601},
  numpages = {6},
  year = {2023},
  month = {Jan},
  publisher = {American Physical Society},
  doi = {10.1103/PhysRevLett.130.033601},
  url = {https://link.aps.org/doi/10.1103/PhysRevLett.130.033601}
}

@article{Metelmann2025Kerr,
  title = {{Kerr-enhanced optomechanical cooling in the unresolved-sideband regime}},
  author = {Diaz-Naufal, N. and Deeg, L. and Zoepfl, D. and Schneider, C. M. F. and Juan, M. L. and Kirchmair, G. and Metelmann, A.},
  journal = {Phys. Rev. A},
  volume = {111},
  issue = {5},
  pages = {053505},
  numpages = {15},
  year = {2025},
  month = {May},
  publisher = {American Physical Society},
  doi = {10.1103/PhysRevA.111.053505},
  url = {https://link.aps.org/doi/10.1103/PhysRevA.111.053505}
}

@article{Qutip1,
title = {{QuTiP: An open-source Python framework for the dynamics of open quantum systems}},
journal = {Comput. Phys. Commun.},
volume = {183},
number = {8},
pages = {1760},
year = {2012},
issn = {0010-4655},
doi = {https://doi.org/10.1016/j.cpc.2012.02.021},
url = {https://www.sciencedirect.com/science/article/pii/S0010465512000835},
author = {J.R. Johansson and P.D. Nation and Franco Nori}
}

@article{Qutip2,
title = {{QuTiP 2: A Python framework for the dynamics of open quantum systems}},
journal = {Comput. Phys. Commun.},
volume = {184},
number = {4},
pages = {1234},
year = {2013},
issn = {0010-4655},
doi = {https://doi.org/10.1016/j.cpc.2012.11.019},
url = {https://www.sciencedirect.com/science/article/pii/S0010465512003955},
author = {J.R. Johansson and P.D. Nation and Franco Nori}
}

@article{VannerTomography2014,
  title = {{Towards optomechanical quantum state reconstruction of mechanical motion}},
  volume = {527},
  ISSN = {1521-3889},
  url = {http://dx.doi.org/10.1002/andp.201400124},
  DOI = {10.1002/andp.201400124},
  number = {1–2},
  journal = {Ann. Phys. (Berl.)},
  publisher = {Wiley},
  author = {Vanner,  Michael R. and Pikovski,  Igor and Kim,  M. S.},
  year = {2014},
  month = aug,
  pages = {15}
}

@article{ShomroniTomography2019,
  title = {{Optical backaction-evading measurement of a mechanical oscillator}},
  volume = {10},
  ISSN = {2041-1723},
  url = {http://dx.doi.org/10.1038/s41467-019-10024-3},
  DOI = {10.1038/s41467-019-10024-3},
  number = {1},
  pages = {2086},
  journal = {Nat. Commun.},
  publisher = {Springer Science and Business Media LLC},
  author = {Shomroni,  Itay and Qiu,  Liu and Malz,  Daniel and Nunnenkamp,  Andreas and Kippenberg,  Tobias J.},
  year = {2019},
  month = may 
}

@article{LiaoTomography2014,
  title = {{Spectrometric reconstruction of mechanical-motional states in optomechanics}},
  author = {Liao, Jie-Qiao and Nori, Franco},
  journal = {Phys. Rev. A},
  volume = {90},
  issue = {2},
  pages = {023851},
  numpages = {9},
  year = {2014},
  month = {Aug},
  publisher = {American Physical Society},
  doi = {10.1103/PhysRevA.90.023851},
  url = {https://link.aps.org/doi/10.1103/PhysRevA.90.023851}
}

@article{Bothner2022Kerr,
  title = {{Four-wave-cooling to the single phonon level in Kerr optomechanics}},
  volume = {5},
  ISSN = {2399-3650},
  url = {http://dx.doi.org/10.1038/s42005-022-00808-3},
  DOI = {10.1038/s42005-022-00808-3},
  number = {1},
  journal = {Commun. Phys.},
  publisher = {Springer Science and Business Media LLC},
  author = {Bothner,  Daniel and Rodrigues,  Ines C. and Steele,  Gary A.},
  year = {2022},
  month = feb,
  pages = {33}
}

@article{MitraOE2024,
  title = {{Narrow-linewidth Fano microcavities with resonant subwavelength grating mirror}},
  volume = {32},
  ISSN = {1094-4087},
  url = {http://dx.doi.org/10.1364/OE.521329},
  DOI = {10.1364/oe.521329},
  number = {9},
  journal = {Opt. Express},
  publisher = {Optica Publishing Group},
  author = {Mitra,  Trishala and Singh,  Gurpreet and Darki,  Ali Akbar and Madsen,  Søren Peder and Dantan,  Aurélien},
  year = {2024},
  month = apr,
  pages = {15667}
}

@article{WLN2018,
  title = {{Resource theory of quantum non-Gaussianity and Wigner negativity}},
  author = {Albarelli, Francesco and Genoni, Marco G. and Paris, Matteo G. A. and Ferraro, Alessandro},
  journal = {Phys. Rev. A},
  volume = {98},
  issue = {5},
  pages = {052350},
  numpages = {17},
  year = {2018},
  month = {Nov},
  publisher = {American Physical Society},
  doi = {10.1103/PhysRevA.98.052350},
  url = {https://link.aps.org/doi/10.1103/PhysRevA.98.052350}
}

@article{FanGuidedResonanceAnalysis2002,
  title = {{Analysis of guided resonances in photonic crystal slabs}},
  author = {Fan, Shanhui and Joannopoulos, J. D.},
  journal = {Phys. Rev. B},
  volume = {65},
  issue = {23},
  pages = {235112},
  numpages = {8},
  year = {2002},
  month = {Jun},
  publisher = {American Physical Society},
  doi = {10.1103/PhysRevB.65.235112},
  url = {https://link.aps.org/doi/10.1103/PhysRevB.65.235112}
}

@article{EnzianOE2023,
  title = {{Phononically shielded photonic-crystal mirror membranes for cavity quantum optomechanics}},
  volume = {31},
  ISSN = {1094-4087},
  url = {http://dx.doi.org/10.1364/OE.484369},
  DOI = {10.1364/oe.484369},
  number = {8},
  journal = {Opt. Express},
  publisher = {Optica Publishing Group},
  author = {Enzian,  Georg and Wang,  Zihua and Simonsen,  Anders and Mathiassen,  Jonas and Vibel,  Toke and Tsaturyan,  Yeghishe and Tagantsev,  Alexander and Schliesser,  Albert and Polzik,  Eugene S.},
  year = {2023},
  month = apr,
  pages = {13040}
}

@article{TsaturyanNNano2017,
  title = {{Ultracoherent nanomechanical resonators via soft clamping and dissipation dilution}},
  volume = {12},
  ISSN = {1748-3395},
  url = {http://dx.doi.org/10.1038/nnano.2017.101},
  DOI = {10.1038/nnano.2017.101},
  number = {8},
  journal = {Nat. Nanotechnol.},
  publisher = {Springer Science and Business Media LLC},
  author = {Tsaturyan,  Y. and Barg,  A. and Polzik,  E. S. and Schliesser,  A.},
  year = {2017},
  month = jun,
  pages = {776}
}

@article{CiersAPL2025,
  title = {{Membrane phononic crystals for high-Qm mechanical defect modes at MHz frequencies in piezoelectric aluminum nitride}},
  volume = {126},
  ISSN = {1077-3118},
  url = {http://dx.doi.org/10.1063/5.0262362},
  DOI = {10.1063/5.0262362},
  number = {25},
  journal = {Appl. Phys. Lett.},
  publisher = {AIP Publishing},
  author = {Ciers,  Anastasiia and Nindito,  Laurentius Radit and Jung,  Alexander and Pfeifer,  Hannes and Dadgar,  Armin and Strittmatter,  André and Wieczorek,  Witlef},
  year = {2025},
  month = jun,
  pages = {253503}
}

@article{SaarinenOptica2023,
  title = {{Laser cooling a membrane-in-the-middle system close to the quantum ground state from room temperature}},
  volume = {10},
  ISSN = {2334-2536},
  url = {http://dx.doi.org/10.1364/OPTICA.468590},
  DOI = {10.1364/optica.468590},
  number = {3},
  journal = {Optica},
  publisher = {Optica Publishing Group},
  author = {Saarinen,  Sampo A. and Kralj,  Nenad and Langman,  Eric C. and Tsaturyan,  Yeghishe and Schliesser,  Albert},
  year = {2023},
  month = mar,
  pages = {364}
}

@article{YuriPRL2018HighQ,
  title = {{Asymmetric Metasurfaces with High-$Q$ Resonances Governed by Bound States in the Continuum}},
  author = {Koshelev, Kirill and Lepeshov, Sergey and Liu, Mingkai and Bogdanov, Andrey and Kivshar, Yuri},
  journal = {Phys. Rev. Lett.},
  volume = {121},
  issue = {19},
  pages = {193903},
  numpages = {6},
  year = {2018},
  month = {Nov},
  publisher = {American Physical Society},
  doi = {10.1103/PhysRevLett.121.193903},
  url = {https://link.aps.org/doi/10.1103/PhysRevLett.121.193903}
}

@article{HuangNC2023HighQ,
  title = {{Ultrahigh-Q guided mode resonances in an All-dielectric metasurface}},
  volume = {14},
  ISSN = {2041-1723},
  url = {http://dx.doi.org/10.1038/s41467-023-39227-5},
  DOI = {10.1038/s41467-023-39227-5},
  number = {1},
  journal = {Nat. Commun.},
  publisher = {Springer Science and Business Media LLC},
  author = {Huang,  Lujun and Jin,  Rong and Zhou,  Chaobiao and Li,  Guanhai and Xu,  Lei and Overvig,  Adam and Deng,  Fu and Chen,  Xiaoshuang and Lu,  Wei and Alù,  Andrea and Miroshnichenko,  Andrey E.},
  year = {2023},
  month = jun,
  pages = {3433}
}

@article{ZhouNL2025HighQ,
  title = {{Ultrahigh-Q Quasi-BICs via Precision-Controlled Asymmetry in Dielectric Metasurfaces}},
  volume = {25},
  ISSN = {1530-6992},
  url = {http://dx.doi.org/10.1021/acs.nanolett.5c00967},
  DOI = {10.1021/acs.nanolett.5c00967},
  number = {14},
  journal = {Nano Lett.},
  publisher = {American Chemical Society (ACS)},
  author = {Zhou,  Chaobiao and Zhou,  Mimi and Fu,  Zhenchu and He,  Haoxuan and Deng,  Zi-Lan and Xiang,  Hong and Chen,  Xiaoshuang and Lu,  Wei and Li,  Guanhai and Han,  Dezhuan},
  year = {2025},
  month = mar,
  pages = {5916}
}

@article{ZhouLPR2023HighR,
  title = {{Cavity Optomechanical Bistability with an Ultrahigh Reflectivity Photonic Crystal Membrane}},
  volume = {17},
  ISSN = {1863-8899},
  url = {http://dx.doi.org/10.1002/lpor.202300008},
  DOI = {10.1002/lpor.202300008},
  number = {10},
  journal = {Laser Photonics Rev.},
  publisher = {Wiley},
  author = {Zhou,  Feng and Bao,  Yiliang and Gorman,  Jason J. and Lawall,  John R.},
  year = {2023},
  month = aug,
  pages = {2300008}
}

@article{RempeOL1992,
  title = {{Measurement of ultralow losses in an optical interferometer}},
  volume = {17},
  ISSN = {1539-4794},
  url = {http://dx.doi.org/10.1364/OL.17.000363},
  DOI = {10.1364/ol.17.000363},
  number = {5},
  journal = {Opt. Lett.},
  publisher = {Optica Publishing Group},
  author = {Rempe,  G. and Lalezari,  R. and Thompson,  R. J. and Kimble,  H. J.},
  year = {1992},
  month = mar,
  pages = {363}
}

@article{FangNP2017,
  title = {{Generalized non-reciprocity in an optomechanical circuit via synthetic magnetism and reservoir engineering}},
  volume = {13},
  ISSN = {1745-2481},
  url = {http://dx.doi.org/10.1038/nphys4009},
  DOI = {10.1038/nphys4009},
  number = {5},
  journal = {Nat. Phys.},
  publisher = {Springer Science and Business Media LLC},
  author = {Fang,  Kejie and Luo,  Jie and Metelmann,  Anja and Matheny,  Matthew H. and Marquardt,  Florian and Clerk,  Aashish A. and Painter,  Oskar},
  year = {2017},
  month = jan,
  pages = {465}
}

@article{BurgwalNC2023,
  title = {{Enhanced nonlinear optomechanics in a coupled-mode photonic crystal device}},
  volume = {14},
  ISSN = {2041-1723},
  url = {http://dx.doi.org/10.1038/s41467-023-37138-z},
  DOI = {10.1038/s41467-023-37138-z},
  number = {1},
  journal = {Nat. Commun.},
  publisher = {Springer Science and Business Media LLC},
  author = {Burgwal,  Roel and Verhagen,  Ewold},
  year = {2023},
  month = mar,
  pages = {1526}
}

@article{YoussefiNature2022,
  title = {{Topological lattices realized in superconducting circuit optomechanics}},
  volume = {612},
  ISSN = {1476-4687},
  url = {http://dx.doi.org/10.1038/s41586-022-05367-9},
  DOI = {10.1038/s41586-022-05367-9},
  number = {7941},
  journal = {Nature},
  publisher = {Springer Science and Business Media LLC},
  author = {Youssefi,  Amir and Kono,  Shingo and Bancora,  Andrea and Chegnizadeh,  Mahdi and Pan,  Jiahe and Vovk,  Tatiana and Kippenberg,  Tobias J.},
  year = {2022},
  month = dec,
  pages = {666}
}

@article{MetelmannPRX2015,
  title = {{Nonreciprocal Photon Transmission and Amplification via Reservoir Engineering}},
  author = {Metelmann, A. and Clerk, A. A.},
  journal = {Phys. Rev. X},
  volume = {5},
  issue = {2},
  pages = {021025},
  numpages = {16},
  year = {2015},
  month = {Jun},
  publisher = {American Physical Society},
  doi = {10.1103/PhysRevX.5.021025},
  url = {https://link.aps.org/doi/10.1103/PhysRevX.5.021025}
}

@article{YinUSC2022,
  title = {{All-optical quantum simulation of ultrastrong optomechanics}},
  author = {Yin, Xian-Li and Zhou, Yue-Hui and Liao, Jie-Qiao},
  journal = {Phys. Rev. A},
  volume = {105},
  issue = {1},
  pages = {013504},
  numpages = {18},
  year = {2022},
  month = {Jan},
  publisher = {American Physical Society},
  doi = {10.1103/PhysRevA.105.013504},
  url = {https://link.aps.org/doi/10.1103/PhysRevA.105.013504}
}

@article{ClerkDissipativeOMcoupling2009,
  title = {{Quantum Noise Interference and Backaction Cooling in Cavity Nanomechanics}},
  author = {Elste, Florian and Girvin, S. M. and Clerk, A. A.},
  journal = {Phys. Rev. Lett.},
  volume = {102},
  issue = {20},
  pages = {207209},
  numpages = {4},
  year = {2009},
  month = {May},
  publisher = {American Physical Society},
  doi = {10.1103/PhysRevLett.102.207209},
  url = {https://link.aps.org/doi/10.1103/PhysRevLett.102.207209}
}

@article{HotCatState2025,
  title = {{Hot Schr\"{o}dinger cat states}},
  volume = {11},
  ISSN = {2375-2548},
  url = {http://dx.doi.org/10.1126/sciadv.adr4492},
  DOI = {10.1126/sciadv.adr4492},
  number = {14},
  journal = {Sci. Adv.},
  publisher = {American Association for the Advancement of Science (AAAS)},
  author = {Yang,  Ian and Agrenius,  Thomas and Usova,  Vasilisa and Romero-Isart,  Oriol and Kirchmair,  Gerhard},
  year = {2025},
  month = apr,
  pages = {eadr4492}
}

@article{Nunnenkamp2012Cooling,
  title = {{Cooling in the single-photon strong-coupling regime of cavity optomechanics}},
  author = {Nunnenkamp, A. and B\o{}rkje, K. and Girvin, S. M.},
  journal = {Phys. Rev. A},
  volume = {85},
  issue = {5},
  pages = {051803},
  numpages = {4},
  year = {2012},
  month = {May},
  publisher = {American Physical Society},
  doi = {10.1103/PhysRevA.85.051803},
  url = {https://link.aps.org/doi/10.1103/PhysRevA.85.051803}
}

@article{TransducerCooling2022,
  title = {{Optomechanical Ground-State Cooling in a Continuous and Efficient Electro-Optic Transducer}},
  author = {Brubaker, B. M. and Kindem, J. M. and Urmey, M. D. and Mittal, S. and Delaney, R. D. and Burns, P. S. and Vissers, M. R. and Lehnert, K. W. and Regal, C. A.},
  journal = {Phys. Rev. X},
  volume = {12},
  issue = {2},
  pages = {021062},
  numpages = {25},
  year = {2022},
  month = {Jun},
  publisher = {American Physical Society},
  doi = {10.1103/PhysRevX.12.021062},
  url = {https://link.aps.org/doi/10.1103/PhysRevX.12.021062}
}

@article{MHzCoolingKippenberg,
  title = {{Room-temperature quantum optomechanics using an ultralow noise cavity}},
  volume = {626},
  ISSN = {1476-4687},
  url = {http://dx.doi.org/10.1038/s41586-023-06997-3},
  DOI = {10.1038/s41586-023-06997-3},
  number = {7999},
  journal = {Nature},
  publisher = {Springer Science and Business Media LLC},
  author = {Huang,  Guanhao and Beccari,  Alberto and Engelsen,  Nils J. and Kippenberg,  Tobias J.},
  year = {2024},
  month = feb,
  pages = {512}
}

@article{Catability2026,
  title = {{Catability as a Metric for Evaluating Superposed Coherent States}},
  author = {Br\"{a}uer, {\v{S}}imon and Provazn\'{\i}k, Jan and Kala, Vojt\v{e}ch and Marek, Petr},
  journal = {Phys. Rev. Lett.},
  volume = {136},
  issue = {9},
  pages = {090205},
  numpages = {6},
  year = {2026},
  month = {Mar},
  publisher = {American Physical Society},
  doi = {10.1103/qfnt-6drh},
  url = {https://link.aps.org/doi/10.1103/qfnt-6drh}
}

\end{document}